\documentclass[%
aip,
amsmath,amssymb,
reprint,%
]{revtex4-1}

\usepackage[utf8]{inputenc}
\usepackage{fullpage,xcolor}
\usepackage{amsmath,amssymb,amsthm,amsopn}
\usepackage[breaklinks=true, colorlinks=true, linkcolor={blue!90!black}, citecolor={blue!90!black}, urlcolor={blue!90!black}]{hyperref}
\usepackage{siunitx}

\usepackage{subcaption}

\usepackage{graphicx}
\usepackage{dcolumn}
\usepackage{bm}

\usepackage[utf8]{inputenc}
\usepackage[T1]{fontenc}
\usepackage{mathptmx}
\usepackage{etoolbox}

\usepackage{orcidlink}

\begin{document}



\preprint{AIP/123-QED}

\title[]{Optimization of transformer ratio and beam loading in a plasma wakefield accelerator with a structure-exploiting algorithm}

\author{Q. Su}%
\email{xpsqq@g.ucla.edu}
\affiliation{Department of Electrical and Computer Engineering, University of California, Los Angeles, Los Angeles, CA 90095,
USA}
\author{J. Larson\orcidlink{0000-0001-9924-2082}}%
\email{jmlarson@anl.gov}
\affiliation{Mathematics and Computer Science Division, Argonne National Laboratory, Lemont, IL 60439, USA}%

\author{T. N. Dalichaouch}
\affiliation{Department of Physics and Astronomy, University of California, Los Angeles, Los Angeles, CA 90095,
USA\looseness=-1}%

\author{F. Li}
\affiliation{Department of Electrical and Computer Engineering, University of California, Los Angeles, Los Angeles, CA 90095, USA}%

\author{W. An}
\affiliation{Department of Astronomy, Beijing Normal University, Beijing 100875, China}%

\author{L. Hildebrand}
\affiliation{Department of Physics and Astronomy, University of California, Los Angeles, Los Angeles, CA 90095, USA\looseness=-1}%

\author{Y. Zhao}
\affiliation{Department of Physics and Astronomy, University of California, Los Angeles, Los Angeles, CA 90095, USA\looseness=-1}%

\author{V. Decyk}
\affiliation{Department of Physics and Astronomy, University of California, Los Angeles, Los Angeles, CA 90095, USA\looseness=-1}%

\author{P. Alves}
\affiliation{Department of Physics and Astronomy, University of California, Los Angeles, Los Angeles, CA 90095, USA\looseness=-1}%

\author{S. M. Wild\orcidlink{0000-0002-6099-2772}}
\affiliation{Mathematics and Computer Science Division, Argonne National Laboratory, Lemont, IL 60439, USA}%

\affiliation{Applied Mathematics and Computational Research Division, Lawrence Berkeley National Laboratory, Berkeley, CA 94720, USA}%

\author{W. B. Mori}
\affiliation{Department of Physics and Astronomy, University of California, Los Angeles, Los Angeles, CA 90095, USA\looseness=-1}%

\affiliation{Department of Electrical and Computer Engineering, University of California, Los Angeles, Los Angeles, CA 90095, USA}%

\date{\today}

\begin{abstract}

Plasma-based acceleration has emerged as a promising  candidate as an accelerator technology for a future linear collider or a next-generation light source. We consider the plasma wakefield accelerator (PWFA) concept where a plasma wave wake is excited by a particle beam and a trailing beam surfs on the wake.  For a linear collider, the energy transfer efficiency from the drive beam to the wake and from the wake to the trailing beam must be large, while the emittance and energy spread of the trailing bunch  must be preserved. One way to simultaneously achieve this when accelerating electrons is to use longitudinally shaped bunches and nonlinear wakes. In the linear regime, there is an analytical formalism to obtain the optimal shapes. In the nonlinear regime, however, the optimal shape of the driver to maximize the energy transfer efficiency cannot be precisely obtained because currently no theory  describes the wake structure and excitation process for all degrees of nonlinearity. In addition, the ion channel radius is not well defined at the front of the wake where the plasma electrons are not fully blown out by the drive beam. We present results using a novel optimization method to effectively determine a current profile for the drive and trailing beam in PWFA that provides low energy spread, low emittance, and high acceleration efficiency. We parameterize the longitudinal beam current profile as a piecewise-linear function and define optimization objectives. For the trailing beam, the algorithm converges quickly to a nearly inverse trapezoidal trailing beam current profile  similar to that predicted by the ultrarelativistic limit of the nonlinear wakefield theory. For the drive beam, the beam profile found by the optimization in the nonlinear regime that maximizes the transformer ratio also resembles that predicted by linear theory. The current profiles found from the optimization method provide higher transformer ratios compared with the linear ramp predicted by the relativistic limit of the nonlinear theory.
\end{abstract}

\maketitle

\section{Introduction}

Plasma wakefield acceleration (PWFA)~\cite{PhysRevLett.54.693}, in which a  particle beam drives a plasma wave wakefield on which a trailing particle beam surfs, has emerged as a promising candidate for future development of more compact and cost-effective advanced lightsources and  
linear colliders. In PWFA, plasma electrons are blown out by the space charge force of the relativistic particle beam and attracted back to axis by plasma ions, creating a close to spherical ion bubble surrounded by a sheath of electrons. Experiments have shown great potential for PWFA to simultaneously achieve high accelerating gradients, high-quality beams, and high energy transfer efficiency~\cite{hogan2000157,hogan2005multi, blumenfeld2007energy, litos2014high, corde2015multi}. In a two-bunch PWFA, the drive beam transfers its energy to the wake, and a second electron or positron bunch (the trailing beam) can be placed at an appropriate distance behind the drive beam to gain energy from the wake. The trailing beam can be externally injected or self-injected. For linear collider applications,  both the energy transfer efficiency from the drive beam ($\eta_w$) to the wake and from the wake to the trailing beam ($\eta_b$) must be high.
The preservation of the trailing beam quality, for example, the energy spread and emittance, during the acceleration is also important.

In linear theory, a figure of merit that helps characterize the energy transfer efficiency is the transformer ratio $R$. It is defined as the ratio between the maximum absolute values of the accelerating field left behind the driver(or experienced by the trailing bunch if loaded) and the decelerating field inside the driver,

\begin{equation}
R =\frac{\max \left|E_{z}\right|^{+}}{\max\left|E_{z}\right|^{-}}.
\end{equation}

The meaning of the transformer ratio can be seen as follows. Some particles in the drive beam will lose all their energy in a pump depletion distance, $L_{pd}=\frac{\gamma_bmc^2}{e\max\left|E_{z}\right|^{-}}$, where $\gamma_b$ is the Lorentz factor of the drive beam. Thus the maximum energy that a particle in the trailing beam can obtain is $e\max\left|E_{z}\right|^{+} L_{pd}=\gamma_b mc^2 R$. 

In order to obtain high acceleration efficiency, it is important to maximize the amount of drive beam energy that is transferred to the wake. This occurs when all the particles in the drive beam decelerate at the same rate; in other words, the drive beam feels a uniform decelerating field.
Otherwise, some particles in the drive beam will deplete their energy while other particles have substantial remaining energy.
A higher transformer ratio can lead to a higher energy gain of the trailing beam for a given drive beam energy over the acceleration distance. However, there is an inherent tradeoff between the maximum loaded charge of the trailing beam and the transformer ratio. Assuming that both the loaded $E_z$ inside the drive beam and the trailing beam are constant.  If 100\% of the drive beam energy were converted into the trailing beam, the trailing beam charge would be smaller that the drive beam charge by a factor proportional to the transformer ratio R. 

It has been shown~\cite{bane1985wake} that the transformer ratio cannot exceed $2$ in 1D linear theory for drive beams with symmetric current profiles (with respect to the beam propagation direction). A higher transformer ratio can be obtained by using an asymmetric current distribution~\cite{PhysRevLett.56.1252}. It is not immediately obvious how the transformer ratio is connected to efficiency. However, linear theory leads to the conclusion that for a bunch with fixed length and fixed charge, the maximum transformer ratio for the unloaded wake occurs when the decelerating field is 
constant~\cite{bane1985collinear}, which is also the condition for transferring all of the drive bean energy into the wake. Thus, in the linear regime  the transformer ratio is also a metric for efficiency.

%
%

The optimal beam profile for energy transfer efficiency and transformer ratio in the linear regime is therefore defined as one that leads to a constant decelerating field. This current profile in 1D was shown to be a delta function precursor followed by a linear ramp. If the decelerating field is parameterized to go from $0$ to a constant as $(1-e^{-\alpha \zeta})E_0$ when $\alpha \to \infty$, then the current profile that achieves this is~\cite{PhysRevLett.56.1252} 
\begin{equation}
\rho(\zeta) \sim -\frac{E_{0}}{4 \pi \alpha}\left[\left(\alpha^{2}+k_{p}^{2}\right) e^{-\alpha \zeta}+k_{p}^{2}(\alpha \zeta-1)\right], \alpha \to \infty ,
\end{equation}
which reduces to a delta function precursor and a linear ramp in the asymptotic limit where $\alpha \to \infty$. Here $\rho$ is defined as the one-dimensional charge density varying in the longitudinal direction $\zeta =  v_b t - z$. In this context $k_p$ is defined as $\omega_p/v_b$, where $\omega_{p}$ is the plasma oscillation frequency and $v_b$ is the beam velocity. In the limit where $\alpha \to \infty$, the transformer ratio is $R=\sqrt{1+k_p^2 L^2}$ and the ratio of the charge in the precursor to that in the linear ramp region is $2/k_p^2 L^2$. For such a current profile, there is $100\%$ energy transfer efficiency from beam to wake (to truly achieve $100\%$, the energy per particle in the precursor should be half that in the wedge-shaped region).







In 3D, a similar analysis can be used because the Green's function for the wake is the response from a delta function charge distribution. In this regime, the decelerating field varies across the beam. Thus, without shaping the driver in the transverse direction, the entire beam cannot slow down together. 

Henceforth, we will use normalized units unless we explicitly assign physical units. Charge is normalized to electron charge $e$; length to $k_p^{-1}$ defined as $c / \omega_{p}$, where $\omega_{p}$ is the plasma oscillation frequency and $c$ the speed of light; charge density to $e n_{p}$; current density to $e n_{p} c$; electric field to $m c \omega_{p} / e$; and potentials to $m c^{2} / e$. We also use the comoving coordinates of a relativistic beam by making a mathematical  transformation from the $(x, y, z, t)$ 
to the $(x, y, \xi = ct-z, s = t)$ variables.

In the 3D blowout (nonlinear) regime there is currently no precise theoretical formalism to obtain the current profile that flattens the decelerating field and to find out whether this also optimizes $R$. Phenomenological models for nonlinear wakefields have assumed physical descriptions where an ion channel is surrounded by one or more electron sheaths~\cite{lu2006nonlinear, dalichaouch2021multi}. However, these models tend to break down at the front of the wake where plasma electrons are not completely blown out and the ion channel has not yet fully formed. Thus, the optimal longitudinal shape for the drive beam in the 3D nonlinear regime has not been well studied.


Much of our understanding of the structure and fields of plasma wakefields in the blowout regime comes from nonlinear theory developed by Lu et al.~\cite{lu2006nonlinear}. In ref.~\cite{lu2006nonlinear}, it was shown that the shape of the ion channel could be completely described by the plasma wake potential $\psi = \phi - A_z$ and the current profiles of the drive and trailing bunches. To use the theory as a predictive tool, Lu et al. introduced a single-sheath model by modeling the plasma source term profile $S = -(\rho-J_z)$ to obtain an expression for the wake potential $\psi = \phi - A_z$,

\begin{equation} \label{eq:poisson_psi}
\nabla_{\perp}^{2} \psi=- \left(\rho-J_{z}\right),
\end{equation}

This wake potential $\psi$ defines the focusing field $\vec{E}_{\perp}+\hat{z} \times \vec{B}_{\perp}=-\nabla_{\perp} \psi$  and the accelerating field $E_{z}=\frac{\partial \psi}{\partial \xi}$ experienced by relativistic beam particles. It has been proven that in the 3D nonlinear regime, the decelerating (and accelerating) fields do not vary across the beam in the transverse direction, so each slice of the drive beam will slow down together~\cite{PhysRevA.44.R6189, lu2006nonlinear}. 

By applying this theory in the very nonlinear (ultrarelativisitic) limit, it was shown that an adiabatically increasing linear current profile would still provide a nearly constant decelerating field~\cite{lu2009high} even in the nonlinear regime. Based on the nonlinear wakefield theory of Lu et al.~\cite{lu2006nonlinear}, the equation for the innermost electron trajectory in the ultrarelativistic limit where the blowout radius $r_{b} \gg 1$ is given by
\begin{equation}\label{eq:rb_lu}
r_{b} \frac{d^{2} r_{b}}{d \xi^{2}}+2\left[\frac{d r_{b}}{d \xi}\right]^{2}+1=\frac{4 \lambda(\xi)}{r_{b}^{2}}.
\end{equation}
Here, $\lambda(\xi) = \int_{0}^{\infty} rn_{b}(r) d r$ represents a normalized charge per unit length for the beam.
By assuming $r_b \gg 1$, the on-axis wake potential $\psi(r,\xi) = \phi-A_{z}$ can be approximated as $\psi(0,\xi) = \frac{r_{b}^{2}(\xi)}{4}$~\cite{lu2006nonlinear}. 

For an adiabatic response, we can assume  $\frac{d^{2} r_{b}}{d \xi^{2}} \ll 1, \frac{d r_{b}}{d \xi} \ll 1$. In this limit, one can predict~\cite{lu2009high} that a constant decelerating field exists within the drive beam when
\begin{equation}\label{eq:lambda_lu}
\lambda(\xi)=\frac{\xi}{L} \Lambda_{0},~\psi(0, \xi) \approx \frac{\xi}{L} \Lambda_{0}, ~\text{and}~ E_{z}(\xi)=\frac{\partial \psi}{\partial \xi} \approx \frac{\Lambda_{0}}{L},
\end{equation}
where $L$ is the beam length and $\Lambda_0$ is the maximum beam current. Within the described approximations, this would be the current profile that provides the highest efficiency in the nonlinear regime. 


As noted above, in the 1D linear regime one can show that the transformer ratio is maximized when the efficiency is also maximized. However, it is not clear how the transfer efficiency and transformer ratio are related in the nonlinear regime. The accelerating field in a nonlinear wake has a deep spike leading to very large peak accelerating fields for electrons that may not be useful. Furthermore, some plasma electrons are blown out with sufficient energy that they are lost, so not all the driver energy goes to the wake.

One can, however, use existing nonlinear theory to obtain a scaling of the transformer ratio with a bunch length for a driver with fixed charge. We define a peak accelerating field that ignores the deep spike.  For $r_{b} \gg 1$, the slope of accelerating field $\partial E / \partial \xi \approx -1 / 2$. Assuming a spherical ion cavity, the maximum useful accelerating field at the rear of the wake can therefore be estimated using $E_{+} \approx \frac{1}{2} r_{\mathrm{max}} \approx \sqrt{\Lambda_0}$, where $r_{max} \approx 2\sqrt{\Lambda_0}$ by assuming the blowout radius $r_b$ reaches its maximum $r_{\mathrm{max}}$ immediately after the drive beam ends~\cite{lu2009high}. From these assumptions it follows that the maximum transformer ratio is $R=|E_{+}| / |E_{-}| \approx \sqrt{\Lambda_0} /\left(\frac{\Lambda_0}{L}\right)=\frac{L}{\sqrt{\Lambda_0}}$. In linear theory $R$ depends only on $L$. In the nonlinear regime, however, $R$  also depends on the peak charge per unit length $\Lambda_0$, which can be written for fixed charge $Q$ as $2 \pi \Lambda_0=2Q/L$. Thus in the nonlinear regime the transformer ratio scales as $R\approx \frac{\sqrt{2 \pi}L^{3/2}}{(2Q)^{1/2}}$.


Just as in the linear regime, one would expected that a precursor could provide the highest efficiencies. A precursor can rapidly increase the wakefield from which the body of the driver can build upon.  However, no theoretical formalism exists for obtaining the response of a precursor and current ramp when the plasma responds nonlinearly. For example, plasma electrons are not fully blown out at the head of the bunch, so it is invalid to extend Eq.~\ref{eq:rb_lu} to include a precursor. Furthermore, studying nonlinear wakes for which the ultrarelativistic limit is not appropriate is more complicated. We note that experimental evidence indicates that high transformer ratios in the nonlinear regime can be obtained for a triangular shape preceded by a precursor and succeeded by a bump at the tail of the beam~\cite{loisch2018observation}.

In order to accurately assess the overall efficiency, it is also important to examine how a trailing beam with charge of interest absorbs the wake energy, which is quantified by $\eta_b$. This is the subject of beam loading. Katsouleas et al.~\cite{Katsouleas:1987yd} showed that linear theory predicts that a trapezoidal current profile can minimize the energy spread (flatten the wake) and emittance growth. However, there is a trade-off between efficiency and the acceleration gradient felt by the loaded bunch. Beam loading in nonlinear wakes was analyzed by Tzoufras et al.~\cite{tzoufras2009beam} using the nonlinear wakefield theory with a single-sheath model~\cite{lu2006nonlinear}, where it was also found that a nearly trapezoidal current that decreases from front to back
is the optimal shape to flatten the wake. In this regime, emittance preservation for beams with finite energy spread can also be achieved through the use of matched beams, and the efficiency can be very high when compared with linear theory. Recently, an improved description for nonlinear wakefields was developed using a multi-sheath model~\cite{dalichaouch2021multi}. The multi-sheath model naturally extends the single-sheath model~\cite{lu2006nonlinear} by including a second plasma sheath that captures regions where the source term, $S = -(\rho-J_z)$, for slices at the rear of the bubble is negative outside the ion channel. The source term  in Eq.~\ref{eq:poisson_psi} is modeled as three regions: an ion channel with radius $r_b(\xi)$ and $S = -1$, an innermost plasma sheath with width $\Delta_1$ and $S \equiv n_1 > 0$, and an outermost plasma sheath of width $\Delta_2$ for which the source term $S \equiv n_2 < 0$. Integrating Eq.~\ref{eq:poisson_psi}, they obtained an expression for the wake potential~\cite{dalichaouch2021multi}
\begin{equation} \label{eq:psi_multi-sheath}
\begin{aligned}
\psi(r, \xi) &=\psi_{0}(\xi)-\frac{r^{2}}{4} \\
&=\frac{r_{b}^{2}(\xi)}{4}\left(1+\beta^{\prime}\right)-\frac{r^{2}}{4},
\end{aligned}
\end{equation}
where $\psi_0(\xi) = \psi(0, \xi)$ is the on-axis wake potential. The function $\beta^{\prime}$ depends on the parameters of the source term profile and is given by
$$
\begin{aligned}
\beta^{\prime} &=2\left(1+n_{1}\right) \ln \left(1+\alpha_{1}\right)-1 \\
&+2 n_{2}\left(1+\alpha_{1}+\alpha_{2}\right)^{2} \ln \left(1+\frac{\alpha_{2}}{1+\alpha_{1}}\right),
\end{aligned}
$$
where $\alpha_{1} \equiv \frac{\Delta_{1}}{r_{b}}$ and $\alpha_{2} \equiv \frac{\Delta_{2}}{r_{b}}$. The parameters $n_1$,$n_2,\Delta_1,$ and $\Delta_2$ are related by the conservation of charge~\cite{lu2006nonlinear}:
\begin{equation}
\label{eq:conscharge}
\int_0^{\infty} (\rho - J_z) rdr = 0. \\
\end{equation} 
The single-sheath expression of $\psi$ can be obtained by setting $n_2 = 0$. Including an additional sheath enables the modeling of negative wake potentials at the rear of the wake, which is important for beam loading and 
self-injection. It was shown that this multi-sheath model provided a more accurate description of the wakefield and could be applied to obtain a beam current with higher acceleration efficiency and lower energy spread compared with the single-sheath model~\cite{dalichaouch2021multi}. 

Advances in computational power and improved algorithmic development in PWFA simulation tools have opened the possibility to directly determine the optimized current profiles of both the drive beam and the trailing beam from simulation. We define the optimization objective to be those that provide the highest efficiency for a given loaded transformer ratio with the lowest energy spread. Previously, evolutionary  algorithms have been applied in accelerator experiments~\cite{hofler2013innovative,mustapha2009optimization}. Recently, a slice-by-slice loading algorithm has been used to optimize beam loading in PWFA~\cite{PhysRevAccelBeams.23.121301}.  Optimization approaches such as Bayesian optimization~\cite{duris2020bayesian,jalas2021bayesian,shalloo2020automation} and neural networks~\cite{PhysRevLett.126.174801} have also aroused great interest in the community of laser-plasma accelerators.


In this paper we develop a specialized numerical optimization routine that seeks
optimal drive beam and trailing beam current profiles in the nonlinear blowout regime. The method is inspired by the nonlinear least-squares 
solver POUNDERs~\cite{SWCHAP14} with modifications to find a current profile that minimizes the deviation of 
the accelerating field or decelerating field, $E_z$, about its mean.
We couple data from the quasi-static particle-in-cell (PIC) code QuickPIC~\cite{huang2006quickpic, AN2013165} to this modified POUNDERs. We parameterize the beam current profile as a piecewise-linear function for which the beam charge is constant. The transverse profile is assumed to be Gaussian with a fixed spot size. 
We also define an optimization objective to quantify variations in the electric field within the beam (either drive or trailing) about an average. The optimization algorithm can work efficiently to minimize the objective function. The procedure also permits including constraints such as the beam length and/or the beam charge. We first use the optimization method to find the optimal profile for the trailing beam subject to constraining the length. This constraint is also imposed by other optimization methods. As mentioned above, the equations for both the single-sheath~\cite{tzoufras2009beam} and multi-sheath models~\cite{dalichaouch2021multi} for nonlinear wakefields~\cite{lu2006nonlinear} can be integrated for a trailing beam inside a nonlinear wake. There is excellent agreement between the predictions from the multi-sheath model for the current profiles that flatten the wakefield and the optimized results from particle-in-cell (OSIRIS and QuickPIC) simulations. The current profiles are nearly inverse trapezoidal in shape. There are slight differences between the algorithm-searched approach current profiles 
and the theoretical predictions for the beam-loading problem, which will be described shortly. The fact that there is such good agreement gives confidence in both the algorithm-searched process and the theory. 

We next use this method to find the optimal drive beam profile that flattens the decelerating field. In this case we constrain the total charge in the beam.  As noted above, it is not straightforward to use existing nonlinear theory because a fully blown-out wake does not exist at the head of the beam. For this problem it is also not possible to use optimization procedures that rely on a slice-by-slice procedure because the value of the objective decelerating field must be known before the optimization for this kind of approach. Interestingly, the algorithm-searched results find current profiles that are nearly identical  to those predicted by 1D linear theory. The optimal current profile has a precursor followed by a triangular shape. We also find that even in the fully nonlinear regime the profiles that provide the most flattened decelerating field also lead to the largest transformer ratios of the current profiles considered for a given fixed total charge and bunch length.

For the cases examined in this paper, we ignore the effects of ion motion on the focusing and accelerating fields as they are expected to be small~\cite{PhysRevLett.118.244801}. Furthermore, the optimization method described here can straightforwardly be applied to cases where ion motion needs to be included.

\section{The optimization algorithm}\label{Sec:method}

\begin{figure*}[!ht]
	\begin{center}

\begin{subfigure}{0.45\textwidth}
    \centering 
    \includegraphics[width=0.95\textwidth, height = 0.7\textwidth]{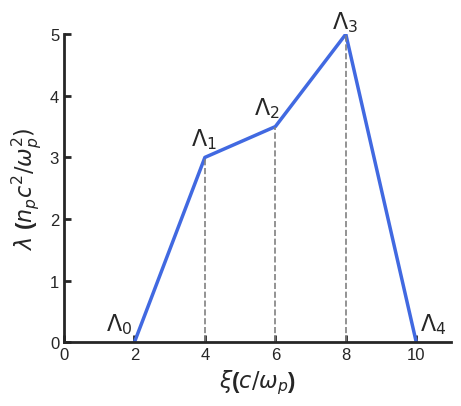}
    \caption{}
\end{subfigure} 
 \begin{subfigure}{0.45\textwidth}
 \centering
 \includegraphics[width=1.05\textwidth, height = 0.7\textwidth]{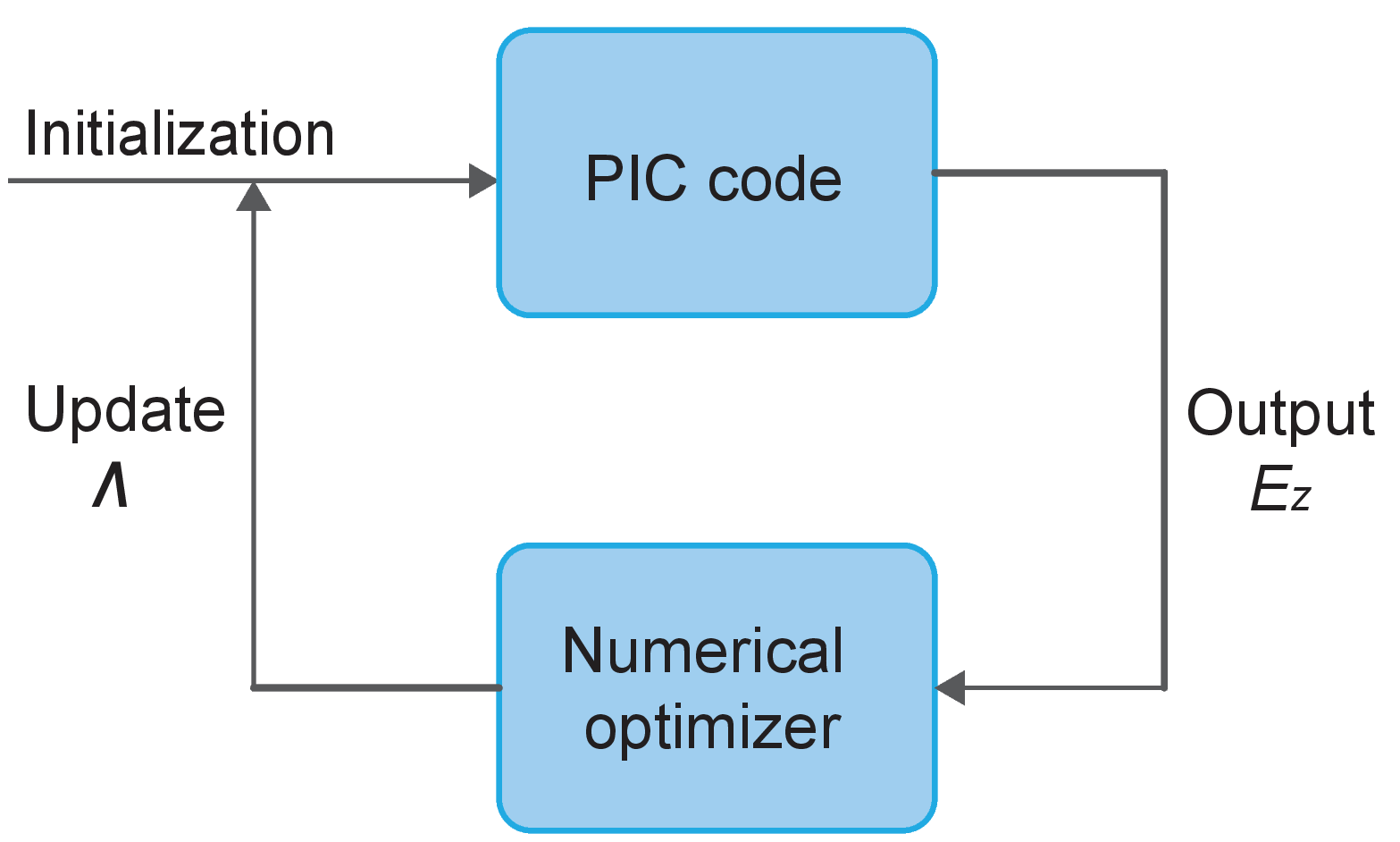}
 \caption{}
 \end{subfigure}  
		
		\caption{(a) Initialization of a beam with a piecewise-linear profile. The optimization variables are set as beam charge (current)  per unit length, $\boldsymbol{\Lambda}$. (b) Optimization workflow. An objective function is evaluated based on the output variables from the PIC simulation. This together with past evaluations is used to determine a new set of values for the optimization variables.   }
            \label{fig:flow}
	\end{center}
\end{figure*}

To optimize the shape of the drive beam, we discretize the beam
longitudinal profile by initializing a piecewise-linear charge density described by the discretized beam charge per unit length $\boldsymbol{\Lambda}$. In QuickPIC, we initialize a longitudinally piecewise-linear drive beam with a
Gaussian transverse profile $n_{b} \sim e^{-r^{2} /\left(2
	\sigma_{r}^{2}\right)}\lambda(\xi)$, where $\lambda(\xi)$ is a piecewise-linear function $h(\xi_i
\leq \xi \leq \xi_{i+1}) = \Lambda_i + \frac{\Lambda_{i+1} -
	\Lambda_i}{\xi_{i+1} - \xi_{i}}(\xi - \xi_i)$ defined by $p$ points of
the discretized normalized charge per unit length $\Lambda_i$ as shown in Fig.~\ref{fig:flow} (a). The positions of the beginning and the end of the beam are fixed, and the currents at these two points are set to zero $\lambda_{\xi_0} = \lambda_{\xi_{n-1}} = 0$.

The objective function is designed as the standard deviation of the field $E_z$. Additional constraints are imposed to fix the total charge $Q$ of the drive beam by $\int \Lambda d\xi =
Q$, or to set upper and lower bounds for the current during the optimization. Under these conditions, the optimization problem can be formulated as
\begin{equation}\label{eq:opt}
\begin{aligned}
& \underset{\Lambda_1,\ldots,\Lambda_p}{\text{minimize}} & & f(\boldsymbol{\Lambda}) = \sum_{k=1}^q ( [\boldsymbol{E}_{z}]_k(\boldsymbol{\Lambda}) - \bar{\boldsymbol{E}}_z(\boldsymbol{\Lambda})  )^2
\\
& \text{subject to:} & &  \sum_{i=1}^{p-1} 0.5  (\Lambda_{i+1} + \Lambda_{i}) \Delta \xi_i = Q,\\
& & & \boldsymbol{\Lambda}_l \le \boldsymbol{\Lambda} \le \boldsymbol{\Lambda}_u,\\
\end{aligned}
\end{equation}
where $Q$ is the total charge and $\Delta \xi_i$ is the piecewise-linear bin
length. The variables that need to be optimized are the beam currents $\Lambda_i$ in each
bin. We set $\Lambda_l = 0$ since the loaded electron beam density should always be greater than 0, and we set $\Lambda_u$ to restrict the search domain to improve the efficiency of the algorithm.  The scalar $q$ counts the total grid points of the discretized vector $\boldsymbol{E_z}$, and $p$ denotes the number of current bins we use. 

The basic concept is to use a quasi-static PIC code to simulate $\boldsymbol{E_z}$ as an unknown function of the normalized beam charge per unit length $\boldsymbol{\Lambda}$ and then use the numerical optimizer to calculate the objective function $f(\boldsymbol{\Lambda})$ and update $\boldsymbol{\Lambda}$. A schematic of this loop is shown in Fig.~\ref{fig:flow} (b). We use QuickPIC to predict the decelerating field or accelerating field $\boldsymbol{E_z}$ in the target beam. For each run, a piecewise-linear beam is initialized by using parameters produced by the
optimization algorithm. Fixing other parameters, QuickPIC gives $E_z$ on the
grid, and the optimization algorithm produces new current profiles. We
formulate the linear constraints as $\sum a_i \Lambda_i = Q$, $a_i =
0.5(\Delta \xi_i + \Delta \xi_{i-1})$ except the first and last components of
$\Lambda_i$, which are set to 0.  $\Lambda_i$ is initialized as $Q
a_i/\|\boldsymbol{a}\|_{2}^2$ to satisfy the linear constraint. 

Solving Eq. \eqref{eq:opt} by applying a general-purpose optimization method will usually 
require passing values of $f$ (as a function of $\boldsymbol{\Lambda}$). While doing so may be easier, it might be inefficient in terms of
the number of times that the simulation must be queried. 
In addition, the objective function $f$ may not be uniquely determined by $\boldsymbol{\Lambda}$, and thus there may exist situations where for different current profiles $\boldsymbol{\Lambda}_1$ and $\boldsymbol{\Lambda}_2$, the values of $f(\boldsymbol{\Lambda}_1)$ and $f(\boldsymbol{\Lambda}_2)$ are equal but the axial fields $E_z(\boldsymbol{\Lambda}_1)$ and
$E_z(\boldsymbol{\Lambda}_2)$ differ considerably. In this situation, an optimization method that requires the user to return only the value of $f$ will likely need additional evaluations to learn about how $f$ varies with $\boldsymbol{\Lambda}$. 
In contrast, the optimization method that can access the information of $E_z(\boldsymbol{\Lambda})$ and regards the objective function $f$ as a function of $E_z$ potentially needs fewer simulation evaluations to minimize $f$.
This is highly desired when each evaluation of $f$
requires a computationally expensive call of QuickPIC to produce
$E_z(\boldsymbol{\Lambda})$.
In fact, our optimization method required only 5 to 10 times the problem dimension $p$ evaluations of QuickPIC to find parameters that produce beams with essentially flat $E_z$ profiles. This is markedly fewer evaluations than required by other commonly employed optimization methods; evolutionary methods such as those used in~\cite{hofler2013innovative, mustapha2009optimization} often require 100 or more objective evaluations in their first generation.

If $\nabla_{\boldsymbol{\Lambda}} E_{z}$ were available, then it would seem
reasonable to search for descent in $f$ in the direction:

\begin{equation}\label{eq:grad}
\begin{aligned}
- \nabla_{\boldsymbol{\Lambda}} f(\boldsymbol{\Lambda}) = & - 2
\sum_k \left[\big([E_{z}]_k
(\boldsymbol{\Lambda}) - \frac{1}{q} \sum_j [E_{z}]_j(\boldsymbol{\Lambda})\big) \right. \\
& \left. \times \big(\nabla_{\boldsymbol{\Lambda}} [E_z]_k(\boldsymbol{\Lambda}) - \frac{1}{q} \sum_j
\nabla_{\boldsymbol{\Lambda}} [E_{z}]_j(\boldsymbol{\Lambda})\big)\right]
\end{aligned} 
\end{equation}
or incorporate second-order knowledge using 
\begin{equation}\label{eq:hessian}
\begin{aligned}
\nabla^2_{\boldsymbol{\Lambda}} f(\boldsymbol{\Lambda}) = &\;  2 \sum_i \left[
\big([E_{z}]_i(\boldsymbol{\Lambda}) - \frac{1}{q} \sum_j
[E_{z}]_j(\boldsymbol{\Lambda})\big) \right. \\
&\left. \times \big(\nabla^2 [E_{z}]_i(\boldsymbol{\Lambda}) -
\frac{1}{q} \sum_j \nabla^2 [E_{z}]_j(\boldsymbol{\Lambda})\big) \right. \\
& \left. + \Big(\nabla [E_{z}]_i(\boldsymbol{\Lambda}) - \frac{1}{q} \sum_j
\nabla [E_{z}]_j(\boldsymbol{\Lambda})\Big) \right. \\
& \left. \times \Big( \nabla
[E_{z}]_i(\boldsymbol{\Lambda}) - \frac{1}{q} \sum_j \nabla [E_{z}]_j(\boldsymbol{\Lambda})\Big)^T\right].
\end{aligned}
\end{equation}
Since gradients of $E_z$ to current profile $\boldsymbol{\Lambda}$ are not available, the modified POUNDERs implementation instead builds local quadratic models
of each of the $q$ mappings $[E_{z}]_k$ around candidate points
$\boldsymbol{\Lambda}_0$. We build these models by interpolating evaluations of
$[E_{z}]_k$ in a neighborhood of $\boldsymbol{\Lambda}_0$. Because each evaluation of QuickPIC for a given set of parameters $\boldsymbol{\Lambda}$
returns all values of $[E_{z}]_k$, the information required to build these $q$ models is easily obtained. 
These approximate gradients and Hessians are used to define a second-order model for the objective; this model is minimized in a neighborhood around the best-known set of parameters $\boldsymbol{\Lambda}$ to generate a candidate point to be evaluated. If this candidate point is an improvement, it becomes the new best-known set of parameters. Otherwise, the previous best-known set of points is kept, and the neighbor size is decreased. In either case, the models of each $[E_z]_i$ can be updated by using any new evaluations. For details, see Ref.~\cite{SWCHAP14}.

\section{Optimization of the trailing beam and comparison with theory}

As a benchmark test, we use our algorithm to obtain the optimal current profile of a trailing beam that flattens the accelerating field of a nonlinear plasma wake. The algorithm results are compared to the optimal profile determined from the nonlinear theory using multi-sheath model (Ref.~\cite{dalichaouch2021multi}). The simulations use normalized units so that each simulation is general and corresponds to a family of different plasma densities. When discussing results for absolute units in  this paper, the plasma densities are assumed to be $n_{p}=1.0 \times 10^{17} ~\si{\cm}^{-3}$, for which $k_p^{-1} = 16.83 ~\si{\um}$.

\subsection{Review of the multi-sheath model}

\begin{figure}[h!]
	\begin{center}
		\includegraphics[width=0.9\linewidth]{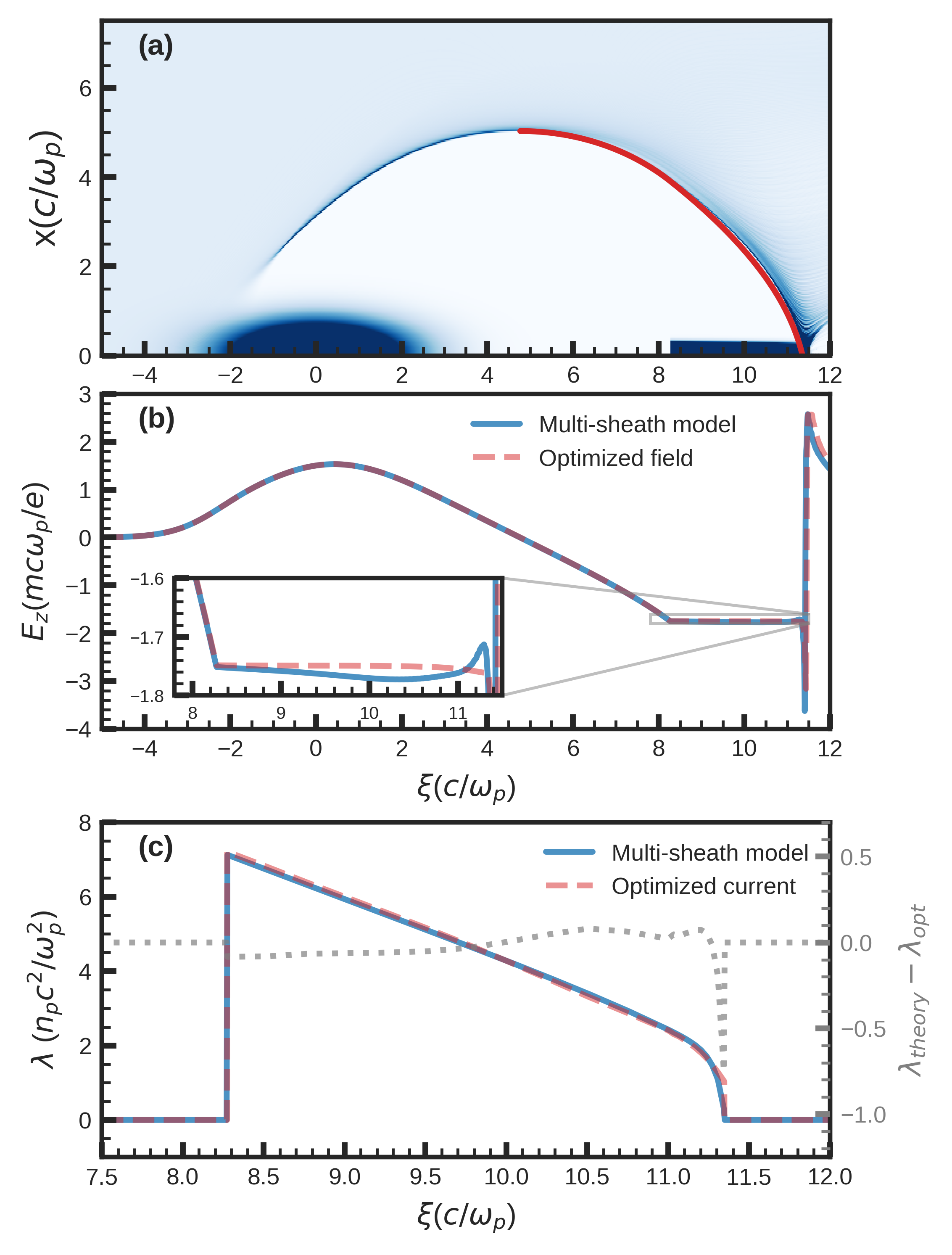}
	\end{center}
	\caption{(a) Two-dimensional cut of the three-dimensional data for the charge density of plasma electrons excited by a bi-Gaussian drive beam and a trailing beam. The charge density of the drive beam and that of the trailing beam are also shown. (b) Comparison of the PIC simulation results for the on-axis wakefield using the optimized profile and the theoretically predicted profiles shown in (c). (c) Comparison of beam profile calculated by the multi-sheath model (blue) and by the optimization procedure using POUNDERs (dashed red). The difference of these profiles is shown as the gray-dotted line. }\label{fig:theory}
\end{figure}

For results obtained using the multi-sheath model, we follow the methodology outlined in Ref.~\cite{dalichaouch2021multi}. We constrain the variables of the multi-sheath model by integrating Eq.~(\ref{eq:conscharge}) to solve for $n_1$ in terms $n_2, \alpha_2,$ and $\alpha_1$:
\begin{equation} \label{eq:const1}
n_{1}=\frac{1-n_{2}\left(\alpha_{2}^{2}+2 \alpha_{2} \alpha_{1}+2 \alpha_{2}\right)}{\left(1+\alpha_{1}\right)^{2}-1}. 
\end{equation}
 The other phenomenological sheath parameters $\Delta_{1}(r_b)$, $\Delta_{2}(r_b)$, and $n_2(r_b)$ are assumed to be functions of the channel radius $r_b(\xi)$ and can be determined empirically from PIC simulations. Here we employ the same profiles used by Dalichaouch et al.~\cite{dalichaouch2021multi} for the first plasma sheath with $\Delta_{1}=\Delta_{10}+\epsilon r_b$ and the second plasma sheath with $n_{2}=n_{20} e^{-s r_{b}^{2} / r_{m}^{2}}$ and $\Delta_{2}=\Delta_{20}$, where $r_m$ is the maximum blowout radius and $n_{20}$ is determined by calculating the limit of $\psi$, which approaches $\psi_{min}$ when $r_b \rightarrow 0$:
\begin{equation}
n_{20}=\frac{2 \psi_{\min }}{\left(\Delta_{10}+\Delta_{20}\right)^{2} \ln \left(1+\frac{\Delta_{20}}{\Delta_{10}}\right)}.
\end{equation}
In Ref.~\cite{dalichaouch2021multi}, it was shown that the differential equation of innermost particle trajectory can be written as 
\begin{equation}\label{eq:rb}
A^{\prime}\left(r_{b}\right) \frac{d^{2} r_{b}}{d \xi^{2}}+B^{\prime}\left(r_{b}\right) r_{b}\left(\frac{d r_{b}}{d \xi}\right)^{2}+C^{\prime}\left(r_{b}\right) r_{b}=\frac{\lambda(\xi)}{r_{b}},
\end{equation}
where the coefficients $A^{\prime}\left(r_{b}\right), B^{\prime}\left(r_{b}\right)$, and $C^{\prime}\left(r_{b}\right)$ are
\begin{equation}\label{eq:rb_thamine}
\begin{aligned}
&A^{\prime}\left(r_{b}\right)=1+\left[\frac{1}{4}+\frac{\beta^{\prime}}{2}+\frac{1}{8} r_{b} \frac{d \beta^{\prime}}{d r_{b}}\right] r_{b}^{2}, \\
&B^{\prime}\left(r_{b}\right)=\frac{1}{2}+\frac{3}{4} \beta^{\prime}+\frac{3}{4} r_{b} \frac{d \beta^{\prime}}{d r_{b}}+\frac{1}{8} r_{b}^{2} \frac{d^{2} \beta^{\prime}}{d r_{b}^{2}}, \\
&C^{\prime}\left(r_{b}\right)=\frac{1}{4}\left[1+\frac{1}{\left(1+\frac{\beta^{\prime} r_{b}^{2}}{4}\right)^{2}}\right] .
\end{aligned}
\end{equation}
By setting $n_2$ = 0, Eq.~\ref{eq:rb} reduces to the equation of $r_b$ trajectory described by the single-sheath model (Ref.~\cite{lu2006nonlinear}), where $\beta^\prime = \beta = \frac{\left(1+\alpha_{1}\right)^{2} \ln \left(1+\alpha_{1}\right)^{2}}{\left(1+\alpha_{1}\right)^{2}-1}-1$. After calculating the trajectory of $r_b$, the longitudinal electric field can be obtained by taking derivatives of $\psi$ at the central axis,
\begin{equation}\label{eq:ez}
E_{z}(\xi)=\frac{d}{d \xi} \psi_{0}(\xi)=D^{\prime}\left(r_{b}\right) r_{b} \frac{d r_{b}}{d \xi},
\end{equation}
where $D^{\prime}\left(r_{b}\right) \equiv \frac{1}{2}+\frac{\beta^{\prime}}{2}+\frac{1}{4} r_{b} \frac{d \beta^{\prime}}{d r_{b}}$. The derivative of $E_z$ in the $\xi$ direction can then be calculated by 
\begin{equation}\label{eq:dez}
\frac{d E_{z}}{d \xi}=D^{\prime}\left(r_{b}\right) r_{b} \frac{d^{2} r_{b}}{d \xi^{2}}+F^{\prime}\left(r_{b}\right)\left(\frac{d r_{b}}{d \xi}\right)^{2},
\end{equation}
where $F^{\prime}\left(r_{b}\right) \equiv D^{\prime}\left(r_{b}\right)+\frac{3}{4} r_{b} \frac{d \beta^{\prime}}{d r_{b}}+\frac{1}{4} r_{b}^{2} \frac{d^{2} \beta^{\prime}}{d r_{b}^{2}}$. In Ref.~\cite{dalichaouch2021multi}, it was shown that Eqs.~(\ref{eq:rb})--(\ref{eq:dez}) can also be used to design a beam current profile that produces a desired wakefield $f(\xi) = E_{z}\left(\xi_{t} \leq \xi \leq \xi_{f}\right)$. For a given function $f(\xi)$, Eq.~(\ref{eq:ez}) can be integrated to obtain the loaded bubble trajectory $\tilde{r}_b(\xi)$. Equations~(\ref{eq:rb})--(\ref{eq:dez}) can then be used to reverse engineer the current profile $\lambda(\xi)$ of the beam load in terms of $\tilde{r}_b(\xi)$ and $f(\xi)$: 
\begin{equation}\label{eq:lambda}
\lambda(\xi)=C^{\prime} \tilde{r}_{b}^{2}+\left(\frac{B^{\prime}}{D^{\prime 2}}-\frac{A^{\prime} F^{\prime}}{D^{\prime 3} \tilde{r}_{b}^{2}}\right) f(\xi)^{2}+\left(\frac{A^{\prime}}{D^{\prime}}\right) \frac{d f(\xi)}{d \xi}.
\end{equation}
To load a constant wakefield, one can simply set $f(\xi) = -E_t$ and $df(\xi)/d\xi = 0$, where  $E_t$ is the desired constant loaded wakefield started at $\xi_t$. We note that the multi-sheath model fundamentally differs from analytic theory (Ref. ~\cite{tzoufras2009beam}), which relies on solving Eq.~\ref{eq:rb_lu} in the ultrarelativistic limit (i.e. $r_b \gg 1$) where all sheath terms are neglected. In the ultrarelativistic limit, it was shown that the wakefield can be flattened by using an trailing beam with a trapezoidal current profile $\lambda(\xi)=\sqrt{E_{t}^{4}+\frac{R_{b}^{4}}{2^{4}}}-E_{t}\left(\xi-\xi_{t}\right)$, where $R_b$ is the maximum blowout radius. To calculate the maximum loaded beam length, we can integrate $\psi$ from $\xi_t$ to $\xi_f$ to get $\psi_{min}-\psi_t = \int_{\xi_{t}}^{\xi_{f}} E_{z} d \xi$, where at $\xi_f$ we assume $r_b = 0$, and $\psi$ becomes $\psi_{min}$. Assuming $r_b \gg 1$ at $\xi_t$ in the ultrarelativistic limit, we can ignore $\beta^\prime$ and obtain $\psi(0,\xi) = r_b^2/4$ from Eq.~\ref{eq:psi_multi-sheath}. This model for the wake potential exhibits the same asymptotic behavior as the single-sheath model~\cite{tzoufras2009beam}, where $\psi_{min} \rightarrow 0$ as $r_b \rightarrow 0$, and the maximum loaded length becomes $\Delta \xi = \frac{r_{t}^{2}}{4 E_{t}}$. On the other hand, the multi-sheath model is valid for all $r_t$ and predicts a longer loaded length $\Delta \xi=\frac{r_{t}^{2}}{4 E_{t}}+\frac{\beta^{\prime}(r_t) r_t^2/4-\psi_{\min } }{E_{t}}$.
\subsection{Comparison of the optimized trailing beam current with the profile obtained from the multi-sheath model}
To check that the optimization procedure agrees with the multi-sheath model, we consider the case studied both by Tzoufras et al.~\cite{tzoufras2009beam} and by Dalichaouch et al.~\cite{dalichaouch2021multi}, in which a nonlinear plasma wakefield is excited by a drive beam with a bi-Gaussian density distribution $n_{b}(r, \xi)=\left\{N_{b} /\left[(2 \pi)^{3 / 2} \sigma_{r}^{2} \sigma_{z}\right] \right\} e^{-r^{2} /\left(2 \sigma_{r}^{2}\right)} e^{-\xi^{2} /\left(2 \sigma_{z}^{2}\right)}$, where $N_b$ is the number of particles in the beam and $n_{b0} \equiv N_{b} /\left[(2 \pi)^{3 / 2} \sigma_{r}^{2} \sigma_{z}\right] $. The normalized beam parameters are $ \sigma_{r}=0.5,~\sigma_{z}=1.414,~N_{b}=139$ (note that $N_b$ is normalized to $n_p/k_p^3$). The normalized charge per unit length is $\lambda=\int dr r n_b = n_{b0}\sigma_r^2 e^{-\xi^2/(2\sigma_z^2)}$. The peak charge per unit length $\lambda$ of the driver is $\Lambda_{0}=6.24$, and drive beam energy is $\gamma_{b}=20000$ with $0$ energy spread. As seen in Fig.~\ref{fig:theory} (a), the drive beam excites a plasma wakefield with maximum blowout radius $r_{m} \simeq 2 \Lambda_{0}^{1 / 2} \simeq  5.0$. We aim to load the trailing beam starting at $\xi_{t}=8.27$, where the bubble radius is $r_t \simeq 3.91$, and the wakefield is $E_{t} \simeq 1.75$. To understand how much charge can be loaded for parameters of interest, we assume a plasma density of $n_{p}=1.0 \times 10^{17} \mathrm{~cm}^{-3}$. For the normalized parameters given above, $k_p^{-1} = 16.83 \mathrm{~\mu m}$,  $\sigma_z= 23.8 \mathrm {~\mu m}$, $\sigma_r= 8.4 \mathrm{~\mu m}$,  the charge of the drive beam is $10.6 \mathrm{~nC}$, the peak current of the drive beam is $53.3 \mathrm{~kA}$, and the loaded wakefield is $E_t = 53.2 \mathrm{~GV/m}$. From  Tzoufras et al.~\cite{tzoufras2009beam} $\Delta \xi_{t r}=\frac{r_{t}^{2}}{4 E_{t}} \simeq 2.18$, while the multi-sheath model~\cite{dalichaouch2021multi} predicts a longer loading length of $\Delta \xi_{t r} \simeq 3.09$ by setting $\psi_{min} = -0.9$

and setting the integration parameters in the multi-sheath model as $\Delta_{1}=0.825+0.05 r_{b}$, $ \Delta_{2}=3$, and $s = 3$.

We obtain the optimized trailing beam profile both from the multi-sheath model and from the optimization algorithm by initializing the trailing beam with a longitudinally piecewise-linear profile from $\xi_t = 8.274$ to $\xi = 11.3498$. For the multi-sheath model, the loaded bubble trajectory is numerically integrated from Eq.~\ref{eq:rb} for the desired electric field profile $f(\xi > \xi_t) = E_t$ and $df/d\xi = 0$. The current profile that produces this wakefield is then calculated directly from Eq.~\ref{eq:lambda}. For the optimization algorithm, we set 22 bins\footnote{While fewer than 10 bins did not fully capture the beam dynamics, increasing/decreasing the number of bins beyond 20 had only a marginal effect on the simulation outputs.} to resolve the profile. We ran QuickPIC for a single 3-D time step to get the acceleration field within the trailing beam. At each iteration within the numerical method, we calculated the optimization objective as Eq.~\ref{eq:opt} and used the method discussed in Sec.~\ref{Sec:method} to optimize the trailing beam profile. To reduce the computational expense, we first set a resolution of $0.11 \times 0.11 \times 0.03 $ in the QuickPIC simulation to obtain a low-fidelity optimization result. We then increased the resolution of the QuickPIC simulation up to as fine as $0.03 \times 0.03 \times 0.009$ and used the previously obtained low-fidelity solution as the initialization to further refine the optimal profile. 

The optimized accelerating field (wakefield) and optimized current profile ($\lambda_{opt}$) are shown in Figs.~\ref{fig:theory} (b) and (c), respectively, as dashed-red lines. For comparison the accelerating field and current profiles predicted from the 
multi-sheath model are shown as solid blue lines. One can also see in Figs.~\ref{fig:theory} (b) and (c) that the on-axis wakefield is flattened at the location of the trailing beam for the profile predicted by multi-sheath model, and the optimized current profile is slightly different from the profile predicted by the multi-sheath model ($\lambda_{ms}$). To make these subtle differences clearer, we also plot $\lambda_{opt}-\lambda_{ms}$ in Fig.~\ref{fig:theory} (c). In the blowout regime the accelerating field does not vary across the transverse cross section of the beams~\cite{PhysRevA.44.R6189, lu2006nonlinear}, so we can assume the whole beam feels the same accelerating field when the on-axis $E_z$ field is flattened. In the inset of Fig.~\ref{fig:theory} (b) we show a zoomed-in plot of the flattened accelerating field showing that the current profiles obtained from the optimization method provide flatter accelerating fields than those obtained from the theoretical framework and that the theoretical framework works well. On this scale, we can see that the relative variation of the 
 $E_z$ field within the trailing beam, defined by $\sigma_{E_z} / \overline{E_z}$, where $\sigma_{E_z}$ is defined as the standard deviation of $E_z$ and $\overline{E_z}$ is defined as the average of the of $E_z$, is less than 1\% in both cases. Analysis of the data shows that  the relative variation of $E_z$ is 0.1\% for the optimized trailing beam profile and 0.6\% for a simulation with the current profile based on the theoretical prediction. Therefore, the optimization algorithm improves the variation of the accelerating field with a small modification to the trailing beam current profile. 
The excellent agreement between the predicted and algorithm-obtained optimal current profiles confirm both the theory and the optimization method. We also  note that the relative differences between the current profiles obtained through optimization and theory may differ slightly depending on strength of the driver ($\Lambda_0$) and the location of the head of the trailing beam.

\begin{figure}[!htbp]
	\begin{center}
		\includegraphics[width=0.9\linewidth]{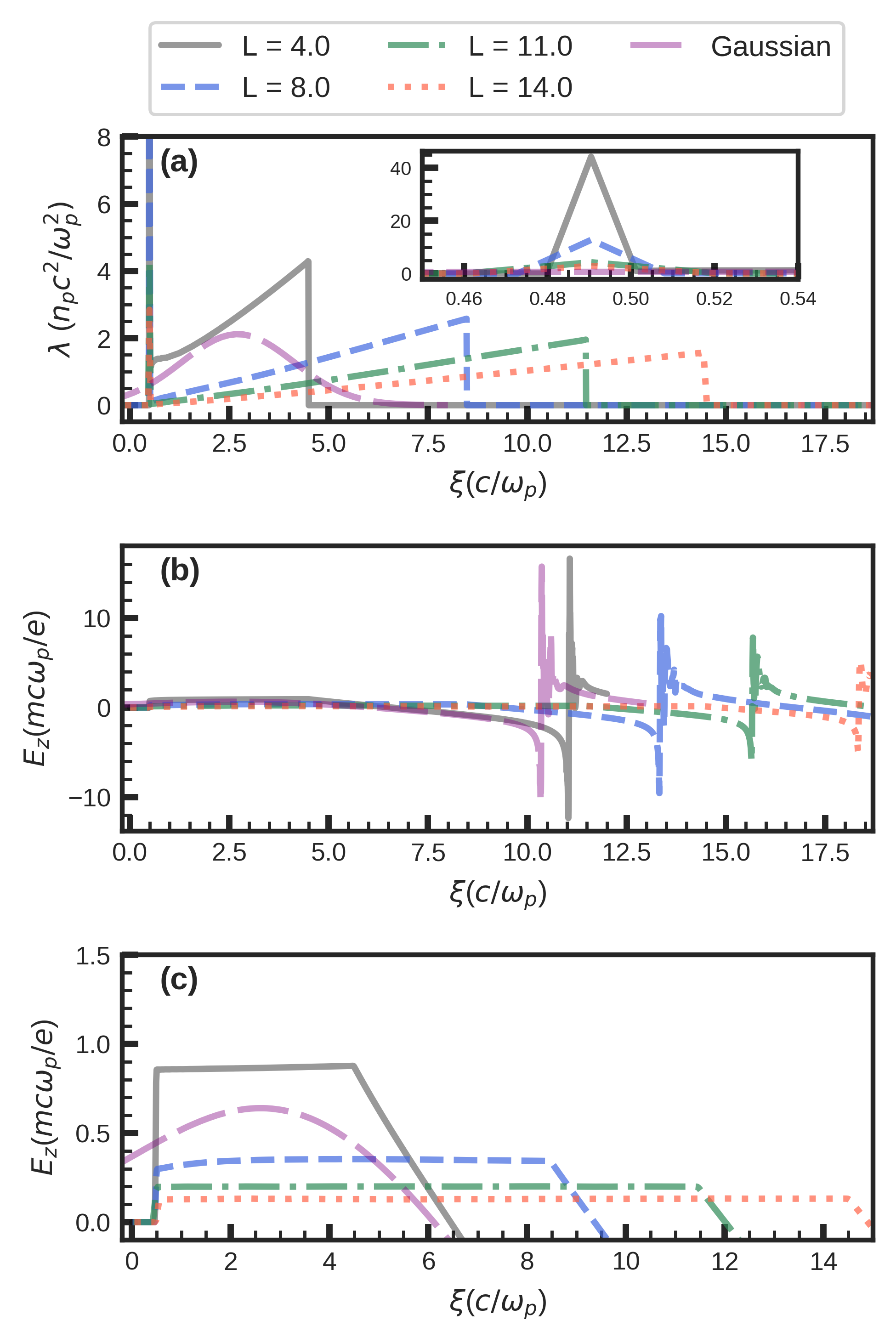}
	\end{center}
	\caption{(a) Optimized current profiles obtained when fixing the total charge of the drive beam as $5 \mathrm{~nC}$, and varying the beam length from 4.0 (gray), 8.0 (dashed blue), 11.0 (dash-dotted green), and 14.0 (dotted red). The bin size $\Delta \xi$ of the piecewise-linear profile is set to the simulation grid size at the beginning and end of the beam and is larger in between. The current profile for a Gaussian with charge of $5 \mathrm{~nC}$ and pulse length of $\sigma = 2$ is also presented (long dashed-purple). (b) On-axis electric field $E_z (\xi)$ using the optimized current profiles and for the Gaussian profile in (a). (c) On-axis electric field $E_z (\xi)$ at the head of the drive beam for each case in (b).}\label{fig:cu_ez_refine}
\end{figure}

\begin{figure}[!htbp]
	\begin{center}
		\includegraphics[width=0.9\linewidth]{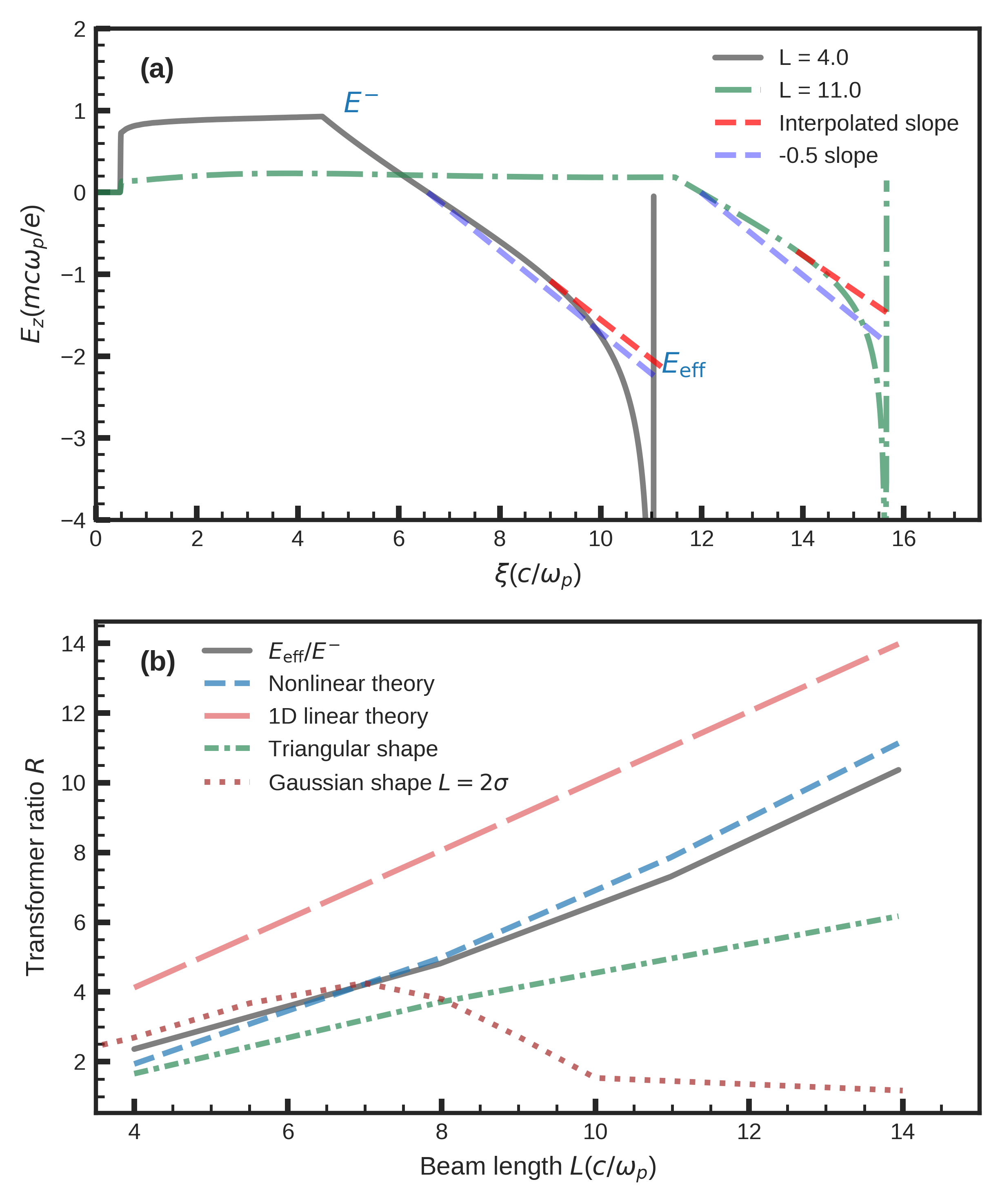}
	\end{center}
	\caption{(a) Optimized on-axis electric field $E_z (\xi)$ for  $L = 4.0$ (gray) and $L = 11.0$ (dash-dotted green). The two methods used to define $E_{\mathrm{eff}}$ and $E^{-}$ are also shown. The dashed blue line starts at $E_z = 0$ at the maximum blowout radius with a slope of $\partial E / \partial \xi=-1 / 2$, and the dashed red line has a slope  extrapolated from the simulated $E_z$ field where it has a nearly linear slope. $E_{\mathrm{eff}}$ is defined as the wakefield where the dashed red line and simulated $E_z$ field meet at the back of the bubble. (b) Simulated transformer ratio $R = E_{\mathrm{eff}}/E^{-}$ as a function of the drive beam lengths $L \in \{4, 8, 11, 14 \}$ for the optimized beam profile (gray), triangular shape longitudinal profile (dash-dotted green), and Gaussian shape (dotted red) with a pulse length $\sigma = L/2$. 
 The theoretical transformer ratio $R = \frac{L}{\sqrt{\Lambda_0}}$ (dashed blue) is calculated by using the value of the current $\Lambda_0$ at the rear of the beam in the simulation; the transformer ratio for linear theory (long-dashed red) is calculated by  $R=\sqrt{(1+ L^2})$.}\label{fig:tr_refine}
\end{figure}

\section{Optimization of the drive beam current profile by fixing the total charge}

\subsection{Optimized driver current profile for different length}
Now that we have verified that the optimization procedure works, we next turn our attention to the drive beam. To optimize the drive beam current profile, we initialize a piecewise-linear monoenergetic drive beam with $\gamma_{b}= 55773, \sigma_{r} = 0.1$. We assume a  fixed total particle number $2 \pi n_0 (k_p^{-1})^3 \int_0^L \lambda(\xi) d\xi$, which corresponds to $N_b= 65.55 \times n_0 (k_p^{-1})^3 = 3.125 \times 10^{10}$ or $5 \mathrm{~nC}$ for a plasma density of $n_{0}=1.0 \times 10^{17} \mathrm{~cm}^{-3}$. Constraining the charge to be fixed is a linear constraint to the current profile of the driver in Eq.~\ref{eq:opt}. Without this constraint, the optimized current tends to converge to all zero values, which is indeed a trivial (but uninteresting) solution for flattening the decelerating field. In this example, we set the transverse resolution to be $0.12 \times 0.12$ in the QuickPIC simulations. To fully resolve the current profile and reduce the number of optimization variables, we use nonuniform bins where the bin size equals the simulation resolution, $\Delta \xi$, at the beginning and the end of the beam, and is set to a larger bin size for the middle of the beam. We also fix the density for the first and last bin to be 0 and examine several cases with different drive beam lengths $L \in \{4, 8, 11, 14 \}$. We choose an axial simulation window size of $5.5$, $8.5$, $11.5$, and $14.5$, respectively, for the drive beam lengths of  $L \in \{4, 8, 11, 14 \}$. The number of grids chosen for each simulation is fixed as  $512 \times 512 \times 512$. The resulting optimized current profiles are shown in Fig.~\ref{fig:cu_ez_refine} (a). When using several bins with a size equal to $\Delta\xi$ at the beginning and end of the beam, we get nearly a ``perfectly" flattened decelerating field, as seen in Fig.~\ref{fig:cu_ez_refine} (b) and (c). Because we are using direct assignment of the piecewise-linear density on the grid when doing simulation (rather than depositing the current from particles), it is possible to have an instantaneous rise of the density in a single grid. Thus the wakefield can directly rise to the  desired constant value at the head of the beam in a length given by the chosen simulation resolution.

We found that in all cases the optimized current profile includes a narrow precursor at the head of the drive bunch. The precursor is always found to have a length of one grid cell no matter the resolution in $\xi$ direction, so one can assume the optimized precursor is a delta function. This was also predicted by 1D linear theory. Although we are operating in a nonlinear regime, an impulse response still is the most efficient method to create the most rapid rise in $E_z$.  Furthermore, as seen in the inset of Fig.~\ref{fig:cu_ez_refine} (a), the amount of charge in the precursor is largest for the shorter beam sizes. This is due to the fact that the decelerating field is smaller for longer beams so the jump in the $E_z$ field required to be generated by the precursor is less. Under the current resolution, we calculated the normalized total charge $Q$ of the optimized precursor when changing the length of the drive beam, then divided the charge $Q$ of each optimized beam by the average decelerating field $E$, where $E/Q$ is close to a constant $68.2 \pm  4.5~\mathrm{GV/(m \cdot nC)}$.

One also can see in Fig.~\ref{fig:cu_ez_refine} (a) that for the shortest beam length, $L=4$ (gray line), the optimal current profile discovered by the algorithm has a gradually increasing slope at the front of the beam (i.e., a nonlinear transition) immediately after the precursor (from $\xi=0.5$ to $\xi=1.5$), which deviates from the prediction of the ultrarelativistic limit of the theory of Lu et al.~\cite{lu2006nonlinear} as described earlier. One can also see in Fig.~\ref{fig:cu_ez_refine} (a) that as the length of the drive bunch is increased while keeping the charge fixed, the profile of the beam after the precursor becomes almost a perfect linear ramp. As seen in Fig.~\ref{fig:cu_ez_refine} (b), both the decelerating and accelerating fields are reduced as the length $L$ is increased. To better visualize the decelerating field, in Fig.~\ref{fig:cu_ez_refine} (c) we plot only the positive region for $E_z$. As can be seen, the decelerating field is nearly flat within the location of the drive beam, and its value decreases as $L$ is increased. 

\begin{figure}[!htbp]
	\begin{center}
		\includegraphics[width=0.9\linewidth]{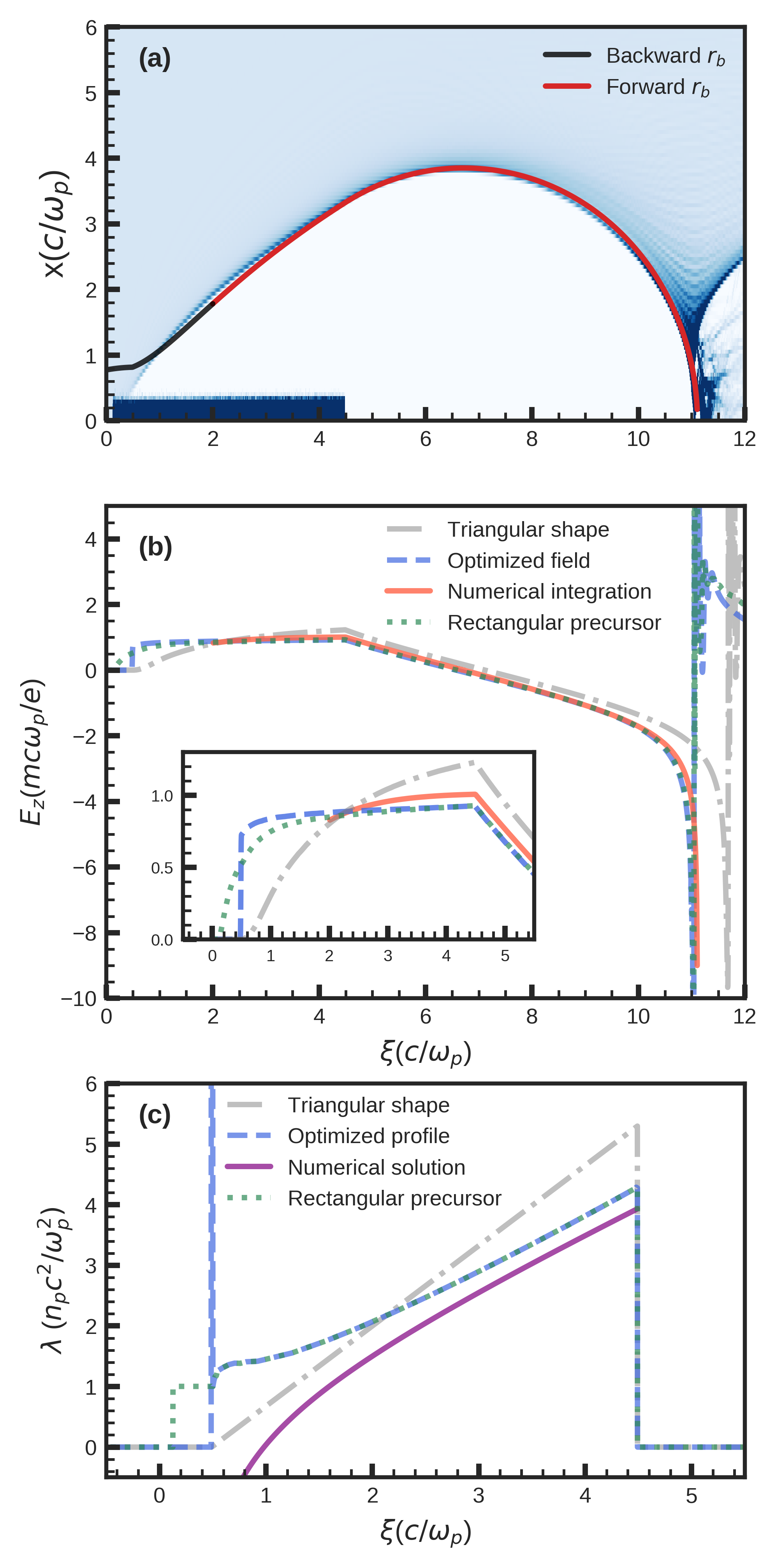}
	\end{center}
	\caption{  (a) Charge density of plasma electrons excited by the optimized drive beam with length $L = 4.0$. The red line represent the bubble radius integrated forward from $\xi = 2.0$ by nonlinear theory including both single-sheath (until $r_{bmax}$ is reached) and multi-sheath (after $r_{bmax}$ is reached) models, and the black line is obtained by integrating $r_b$ backward from $\xi = 2.0$ with the single-sheath model. (b) On-axis accelerating field $E_z(\xi)$ for a triangular profile (dash-dotted gray), optimized profile (dashed blue), rectangular precursor extension of optimized profile (dotted green). The red line represent the predicted $E_z$ field for $\xi > 2$ from Eq.~\ref{eq:ez} given the predicted $r_b$ shown as the red line in (a). (c) Profiles that produce $E_z(\xi)$ shown in (b), and a profile (purple line) calculated from Eq.~\ref{eq:lambda} with the bubble radius $r_t$ and a constant decelerating field $f(\xi) = E_t$ at the $\xi_t$ where the beam ends.}\label{fig:precursor_compare}
\end{figure}

\subsection{Transformer ratios of optimized driver current profiles for different beam lengths}
In nonlinear wakes there is a large negative electron density and a wakefield spike at the rear of the first bubble. Since this spike occurs in a very small region, the peak accelerating field is generally not a good figure of merit when characterizing the transformer ratio. We therefore define the effective maximum wakefield $E_{\mathrm{eff}}$ by extrapolating the part of the wakefield $E_z$ to the rear of the wake (Fig.~\ref{fig:tr_refine} (a)) assuming that it has a slope approximately given by $\partial E / \partial \xi=-1 / 2$. We then define the transformer ratio as the ratio of the effective wakefield maximum $E_{\mathrm{eff}}$ to the maximum decelerating field felt by the drive beam, $R = E_{\mathrm{eff}}/E^{-}$. In Fig.~\ref{fig:tr_refine} (b) we plot the transformer ratio from simulations with the optimized beam profile (gray), a triangular-shaped 
profile (dash-dotted green),  and Gaussian shape (dotted red) for the four different bunch lengths $L \in \{ 4, 8, 11, 14 \}$. For the Gaussian profiles we use $L=2\sigma_z$. The transformer ratio $R$ of the optimized current profile obtained from simulation is also compared with the linear theory using $R=\sqrt{(1+ L^2})$ (long-dashed line), and with the nonlinear theory (dashed blue line) using $E_{+} / E_{-} \approx \sqrt{\Lambda_0} /\left(\frac{\Lambda_0}{L}\right)=\frac{L}{\sqrt{\Lambda_0}}$, where $\Lambda_0$ is the normalized charge per unit length at the end of the optimized beam. We see in Fig.~\ref{fig:tr_refine} (a) that for longer beam lengths, both the peak decelerating field and effective accelerating field become smaller for a decreased peak current. As can be seen in Fig.~\ref{fig:tr_refine} (b)  the transformer ratio increases with bunch length. Thus the accelerating field decreases at a lower rate than does the decelerating field.

\begin{figure}[!htbp]
	\begin{center}
		\includegraphics[width=0.9\linewidth]{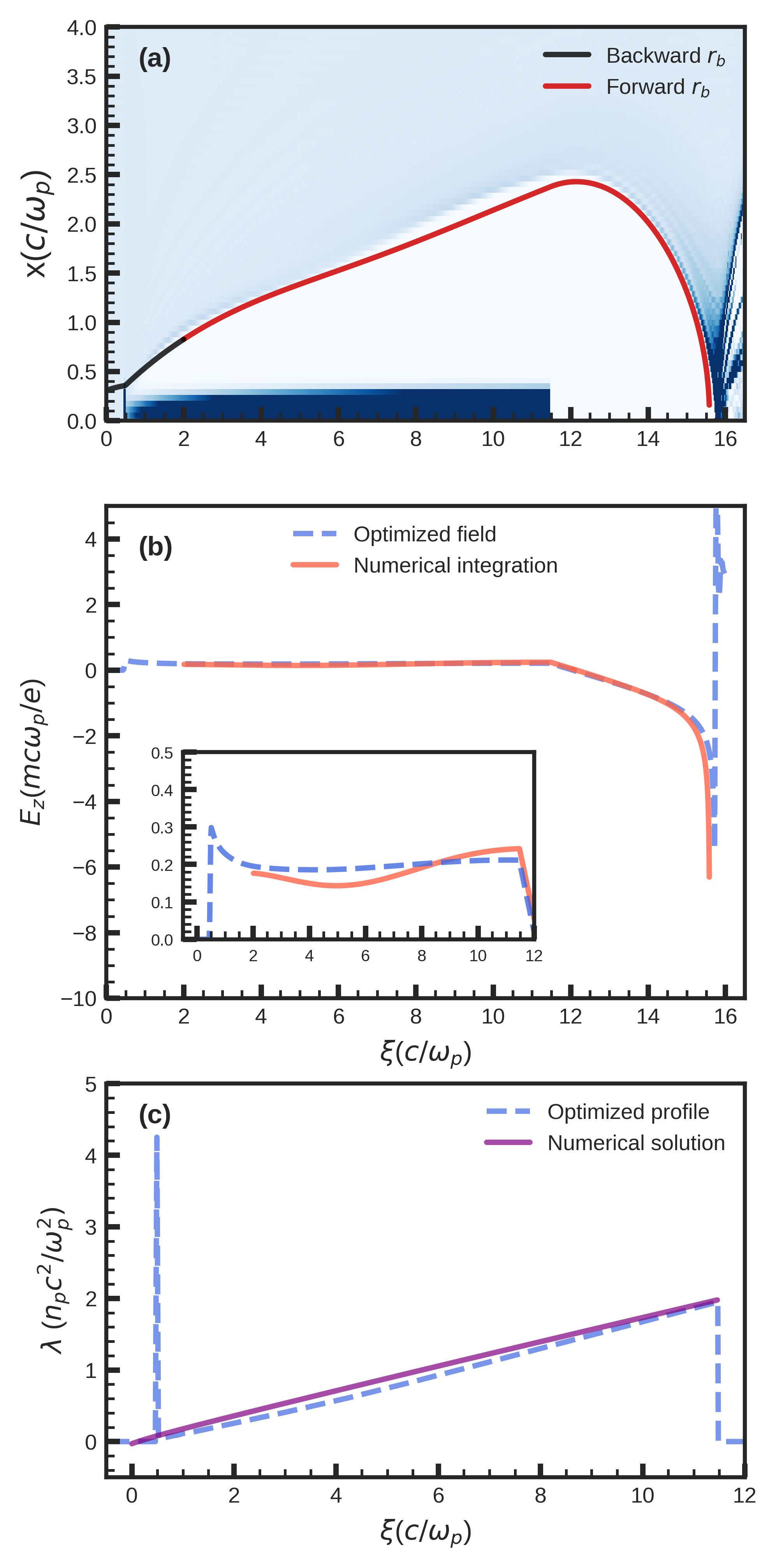}
	\end{center}
	\caption{(a) Charge density of plasma electrons excited by the optimized drive beam with length $L = 11.0$. The red line represents the bubble radius integrated forward (increasing $\xi$)  from $\xi = 2.0$ by the nonlinear theory including both single-sheath and multi-sheath models, and the black line is obtained by integrating $r_b$ backward from $\xi = 2.0$ with the single-sheath model. (b) On-axis accelerating field $E_z(\xi)$ for the optimized trailing beam profile (dash-dotted blue). The red line represent the predicted $E_z$ field for $\xi > 2$ from Eq.~\ref{eq:ez} given the predicted $r_b$ shown as the red line in (a). (c) Profiles that produce the $E_z(\xi)$ shown in (b), and a profile (purple line) calculated from Eq.~\ref{eq:lambda} with the bubble radius $r_t$ and a constant decelerating field $f(\xi) = E_t$ at the $\xi_t$ where the beam ends.}\label{fig:l11}
\end{figure}

To understand the reason that the optimized transformer ratio deviates from the theoretical estimate, we show in Fig.~\ref{fig:precursor_compare} and Fig.~\ref{fig:l11} more detailed analysis for the two cases that have lengths $L = 4$ and $L = 11$. The equation for evolution of the blowout radius $r_b$, Eq.~\ref{eq:rb}, is integrated both forward and backward in $\xi$  for both bunch lengths starting from the initial conditions at $\xi = 2.0$.  The theoretical trajectories for $r_b$ are shown in Fig.~\ref{fig:precursor_compare} (a) and Fig.~\ref{fig:l11} (a) where the red line corresponds to integrating forward to a larger $\xi$ and the black line corresponds to integrating backward in $\xi$. 
We use the single-sheath model for values of $\xi$ before $r_b$ reaches its maximum value, $r_{\mathrm{max}}$. Because the wake potential for the first half bubble is positive definite so only one sheath is required. We then use the multi-sheath model for $\xi$ after $r_b$ reaches its maximum for an accurate description of the negative wake potential at back of the bubble. For both models, we set $\Delta_{1}= 0.825 +0.05 r_{b}$ for $L = 4.0$, $\Delta_{1}= 1.0 +0.1 r_{b}$ for $L = 11.0$, respectively. We set $ \Delta_{2}=3$ and $s = 3$ when using the multi-sheath model.
The theoretically obtained $r_b$ agrees well with the QuickPIC simulation, and the predicted $E_z$ field for $\xi > 2$ from Eq.~\ref{eq:ez} given the predicted $r_b$ also agrees well except for some small oscillations on a smaller scale (red line in Fig.~\ref{fig:precursor_compare} (b) and Fig.~\ref{fig:l11} (b)).

For $L = 4$, the transformer ratio for the optimized current is higher than that estimated from theory. The difference arises because the estimation of Lu et al. for $r_{\mathrm{max}}$ is based on assuming an adiabatic response for $r_b$ from which it follows that the maximum $E_z$ will be reached when the current profile has reached its maximum (at the end of the drive beam). As can be seen in Fig.~\ref{fig:precursor_compare} (a), however, for the $L=4$ case, $r_b$ continues to increase after the drive beam so that  $r_{\mathrm{max}}$ is higher than expected, resulting in a higher effective wakefield $E_{\mathrm{eff}}$ than expected. Furthermore, the optimized profile also gives a lower decelerating field $E_{-}$ (Fig.~\ref{fig:precursor_compare} (c)). These two factors contribute to the slightly higher transformer ratio obtained from optimization. 

For longer pulse $L = 11.0$ (Fig.~\ref{fig:l11}), the optimized transformer ratio is lower than the theoretical estimates. This occurs because as $\Lambda_{0}$ of the drive beam decreases $r_{\mathrm{max}}$ also decreases, and the contribution of $\beta^{\prime}$ can no longer be neglected. In this case, the slope of $E_z$ deviates from the nonlinear limit $\partial E / \partial \xi=-1 / 2$ that is used to estimate $E_{\mathrm{eff}}$ in Fig.~\ref{fig:tr_refine}.

\subsection{Discussion of why nonlinear theory cannot get the precursor}
In reality, we cannot create a precursor with an infinitesimal length, nor can we ramp the drive beam current directly from a peak value to 0 in an infinitesimal distance. Thus, it is useful to  consider constraints on  the designs where the beam parameters are more realizable. We consider the $L=4$ case and constrain the precursor into a rectangular shape (referred to as a doorstep~\cite{bane1985collinear}) with a length of $0.37$ (dotted green line in Fig.~\ref{fig:precursor_compare} (c)) and with same charge as in the ``delta" function precursor (see the gray line in the inset of Fig.~\ref{fig:cu_ez_refine} (a)). The new shape has a smoother transition at the beginning of the beam, as shown in Fig.~\ref{fig:precursor_compare} (c). The transformer ratio does not change; and, as seen in Fig.~\ref{fig:precursor_compare} (b), the decelerating field of the drive beam is nearly constant except for a smooth transition region of length of approximately $k_p^{-1}$ at the beginning of the beam. For comparison, the results for a linear triangular beam current (with the same total charge) are shown (dash-dotted gray line in Fig.~\ref{fig:precursor_compare} (c)), which is the optimal shape predicted from ultrarelativistic nonlinear theory ~\cite{lu2009high}. 
As we have mentioned, we integrate $r_b$ from $\xi = 2.0$ both forward to a larger $\xi$ and backward to the front with Eq.~\ref{eq:rb_thamine} by setting $\beta^\prime = \beta$ and using optimized beam current $\lambda(\xi)$. In Fig.~\ref{fig:precursor_compare} (a)  we can see that the integrated $r_b$ trajectory agrees well with the simulation result when moving forward to the rear of the bubble. On the other hand, when moving backward to the front of the wake, the prediction of the single-sheath model deviates from the simulation result. The lack of agreement in this region arises because there is not a well defined bubble radius at the front of the bubble because of particle crossing, so that the $r_b$ trajectory we get from the nonlinear theory is not the innermost electron sheath near the front of the bubble. This explains why the numerical solution for $\lambda(\xi)$ with Eq.~\ref{eq:lambda} does not recover the precursor (purple line in Fig.~\ref{fig:precursor_compare} (c) and Fig.~\ref{fig:l11} (c)). The same comparison of the optimized profile between theory and simulation are shown in  Fig.~\ref{fig:l11} (c) for the $L=11$ case. The agreement is better because the precursor has less charge.

\subsection{Optimized current profiles for larger bin sizes}
\begin{figure}[!htbp]
	\begin{center}
		\includegraphics[width=0.9\linewidth]{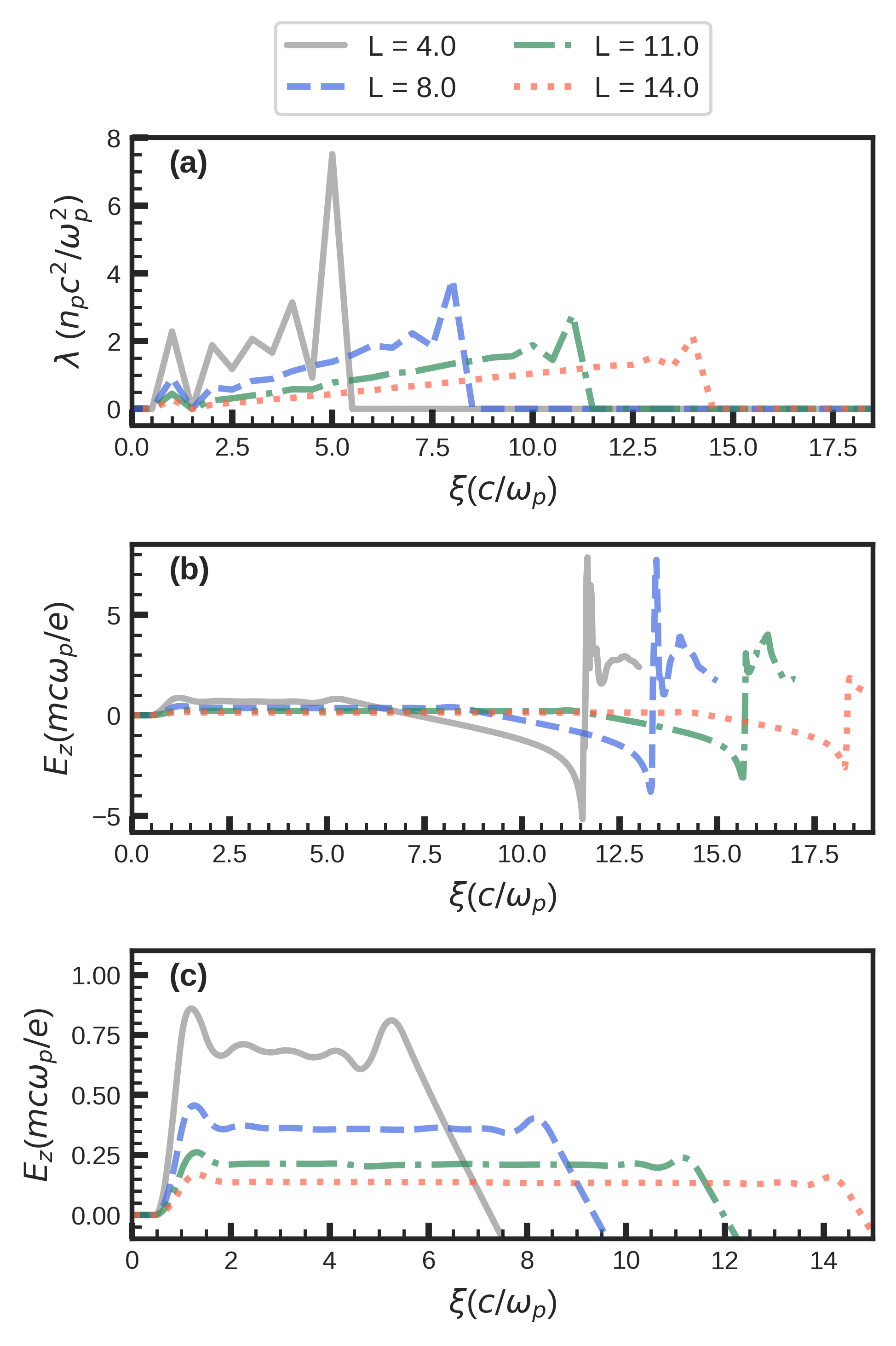}
	\end{center}
	\caption{(a) Optimized current profiles for larger bin sizes when fixing the total charge of drive beam as $5 \mathrm{~nC}$ and varying the beam length over 4.0 (gray), 8.0 (dashed blue), 11.0 (dash-dotted green), and 14.0 (dotted red). Here $k_{p}\Delta \xi$ is fixed to be 0.5. (b) On-axis electric field $E_z (\xi)$ using the  optimized current profile in (a). (c) On-axis electric field $E_z (\xi)$ at the head of the drive beam. }\label{fig:cu_ez}
\end{figure}

When generating current profiles in experiments,  providing precise control of the current profile is difficult. We therefore investigate how the optimized profiles change when the bin sizes are increased. Fig.~\ref{fig:cu_ez} presents results for the varying bunch lengths $L \in \{4, 8, 11, 14 \}$ where the bin sizes are now uniformly set to be $\Delta \xi = 0.5$, while other parameters are kept the same as in Fig.~\ref{fig:cu_ez_refine}. In each case the optimized current profile includes  a precursor with a length of $\Delta \xi$ followed by a quasi-linear ramp, which resembles the solution in Fig.~\ref{fig:cu_ez_refine}. For the larger bin sizes, however, the ``linear" ramp now has oscillations that get smaller for the large bunch sizes.  
For each case shown in Fig.~\ref{fig:cu_ez} (c), the wakefield now smoothly ramps up at the head of the drive beam and then overshoots the average decelerating field because of the finite width of the precursor. The small oscillation in the current profile at the rear of each beam is now present to compensate for the fact that the beam current cannot fall sharply to 0. The optimal shapes found for the coarser bin sizes are close to the beam current profiles used in an experiment that achieved a high transformer ratio~\cite{loisch2018observation}.

\section{A long-range simulation using the optimized drive beam and trailing beam}
\begin{figure}[!htbp]
	\begin{center}
		\includegraphics[width=0.9\linewidth]{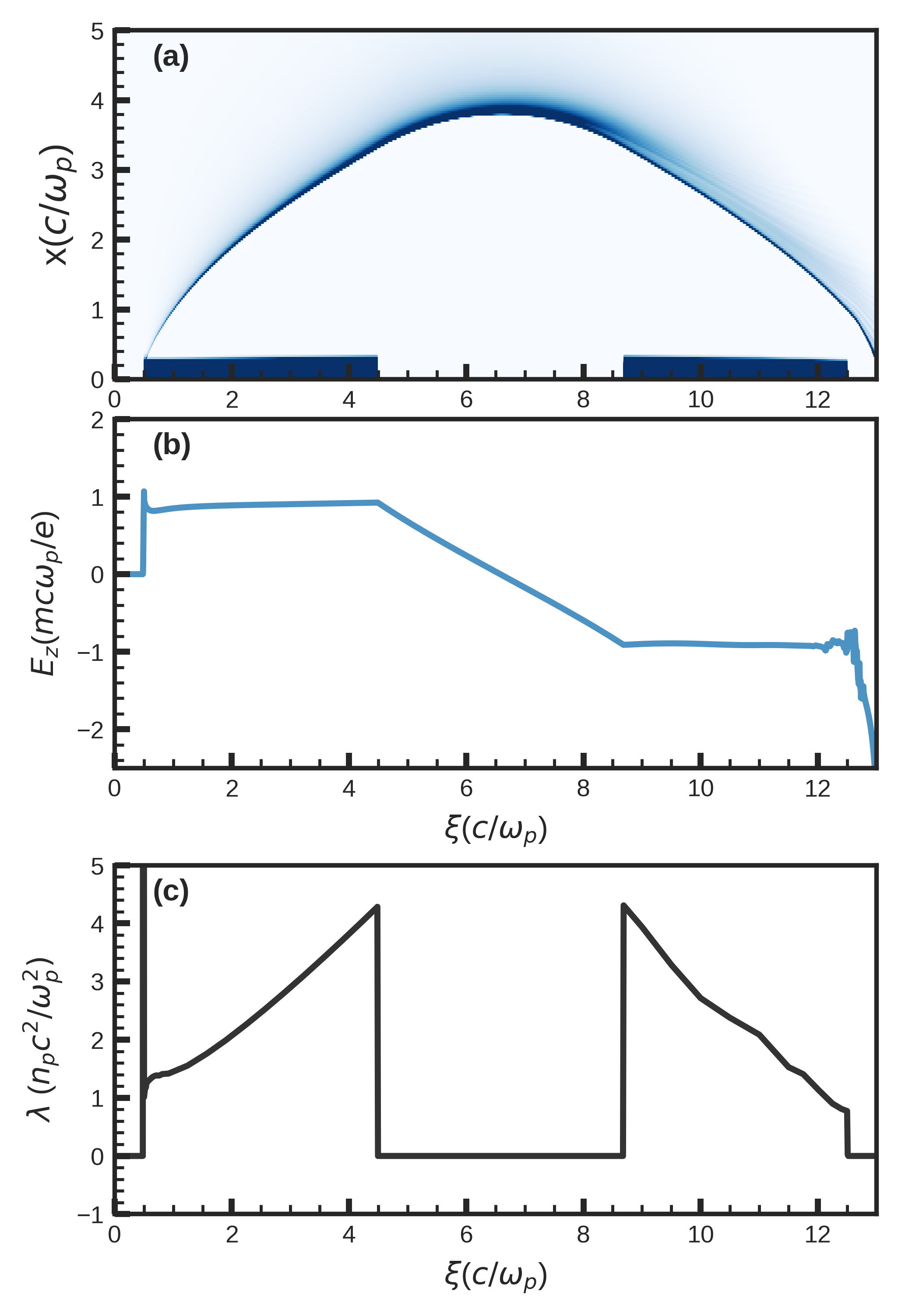}
	\end{center}
	\caption{(a) Charge density of plasma electrons excited by the optimized drive beam and the optimized trailing beam. (b) On-axis accelerating field $E_z(\xi)$ given the optimized current profile. (c) The optimized beam density per unit length $\lambda(\xi)$ for the drive and trailing beams.}\label{fig:opt_wake}
\end{figure}

To confirm that the optimized drive and trailing beam profiles can be used to sustain efficient acceleration over pump depletion distances, we use QPAD,~\cite{li2021quasi}, a highly efficient quasi-static code based on the azimuthal Fourier decomposition method. We initialize the plasma density as $n_{0}=1.0 \times 10^{17} \mathrm{~cm}^{-3}$, and $k_{p}^{-1}=16.83 \mathrm{~\mu m}$. To prevent head erosion and also to ensure that the plasma electrons are fully blown out, we initialize the precursor with a normalized emittance of $0.0167$ and initialize the main body of the driver with a normalized emittance $1.67$. The trailing beam also has a normalized emittance of $1.67$. Both the drive and trailing beams are initialized with $\gamma = 55773$  and matched spot size $\sigma_{r}=\left(2 \epsilon_{N}^{2} / \gamma\right)^{1 / 4}$. We use a drive beam with $L=4.0$ and an optimized current profile (from Fig.~\ref{fig:cu_ez_refine}) and load a trailing beam at $x_t = 8.68$ and $E_t = 0.91$ based on the optimization algorithm (see Fig.~\ref{fig:opt_wake} (a)). We note that at this location the bubble radius  is $r_t = 3.31$. The  decelerating field on the drive beam is flattened to $0.91$, so the loaded transformer ratio is $R \simeq 1.0 $. From the analytic theory shown by Tzoufras et al., the maximum loaded beam length is $\Delta \xi_{t r}=\frac{r_{t}^{2}}{4 E_{t}} \simeq 3.0$. The multi-sheath model predicts a much longer loaded beam length $\Delta \xi_{t r}= \frac{r_{t}^{2}}{4 E_{t}}+\frac{\beta^{\prime}(r_t) r_t^2/4-\psi_{\min } }{E_{t}} \simeq 4.8$ assuming $\psi_{min} = -1$. We initialize 21 bins of a piecewise-linear profile and load a trailing beam with a length of $3.8$. The current profile is then obtained by the described optimization method. The charge in the trailing beam is $4.3 \mathrm{nC}$, which corresponds to  $86\%$ of the drive beam charge.  In the simulation, we use a fine resolution $dr = 0.015$ and $d\xi = 0.0058$ to fully resolve the matched spot size and very short duration of the low emittance precursor. The optimized wakefield is shown in Fig.~\ref{fig:opt_wake} (b).

\begin{figure}[!ht]
	\begin{center}
		\includegraphics[width=0.9\linewidth]{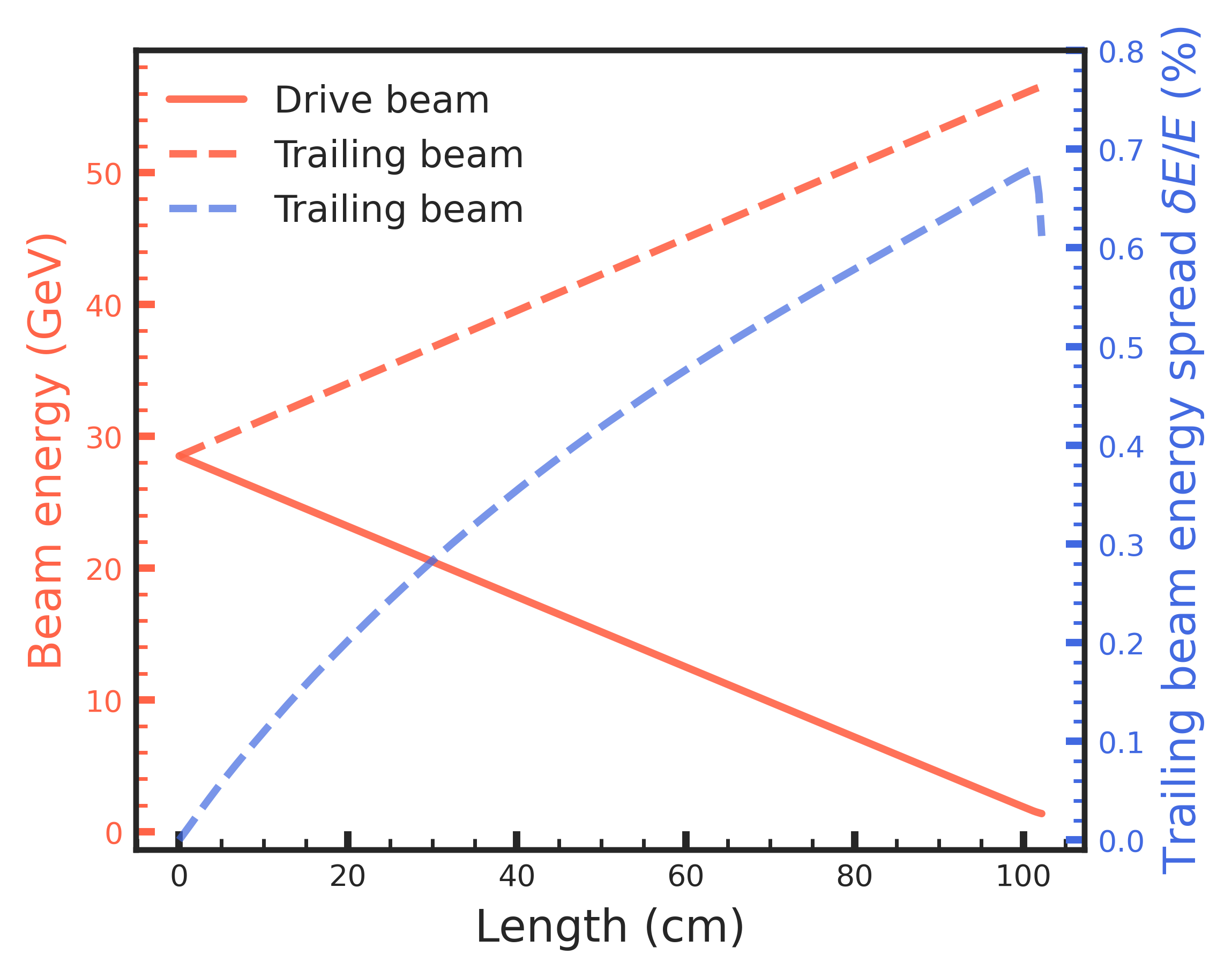}
	\end{center}
	\caption{Drive beam energy depletion (red) and trailing beam energy growth (red dashed). The growth of the energy spread of the trailing beam is shown in blue dashed line.} \label{fig:long_run_energy}
\end{figure}

 We initialize the energy spread of both the drive beam and the trailing beam to 0 to simplify the analysis. The trailing beam is accelerated for $1~\si{\m}$, at which point some drive beam particles have nearly pump depleted to energies around several $\mathrm{MeV}$. The averaged particle energy at different slices of the drive beam and trailing beam is shown in Fig.~\ref{fig:long_run_energy}. The energy of the trailing beam nearly doubles and sustains a stable acceleration with acceleration efficiency $84\%$ from the drive beam to the trailing beam.  The projected energy spread grows from 0 to less than 0.7\%. 


\section{Conclusion}
PWFA has emerged as a promising candidate for the accelerator technology for a future linear collider and/or light source. For the linear collider application, the energy transfer efficiency from the drive beam to the wake and from the wake to the trailing beam must be efficient, and the energy spread of the trailing bunch should be kept low. One way to achieve this is to use longitudinally shaped bunches. In the linear regime, there is an analytical formalism to obtain optimal shapes, for which the transformer ratio is maximized. In the nonlinear blowout regime, however, the theoretical framework is not as well defined for the driver. We thus use a novel optimization method to efficiently find optimized drive beam profiles and trailing beam profiles for PWFA. We parameterize the beam currents as a piecewise-linear longitudinal profile with $N$ bins and define optimization objectives. We use the particle-in-cell code QuickPIC to evaluate the objective, and we use a modified version of POUNDERs (a derivative-free optimization method) to determine a new longitudinal profile. The optimization method required very few evaluations of QuickPIC to identify the best set of parameters. The algorithm is shown to converge quickly and finds trailing beam shapes that are similar to those calculated by a recent multi-sheath model for the nonlinear wakefield. All results were produced when limiting the optimization method objective evaluations to 10 times N, the number of bins.  We also found that even in the nonlinear regime, current profiles for a fixed charge that optimized the efficiency also provided the highest transformer ratio.  In this study, we keep the ions fixed, which will not be the case for the very narrow matched spot sizes of trailing beams for linear collider parameters~\cite{PhysRevLett.118.244801}. The optimization procedure described here should also work well for finding the optimal beam shapes, including the transverse shape, when ion motion is included. There are numerous other applications for the described method within the field of plasma-based acceleration. These include finding optimal shapes for positron acceleration (other optimization methods have already been applied to aspects for this problem~\cite{PhysRevAccelBeams.23.121301}), finding optimal density profiles for matching sections, and finding optimal spatial-temporal couplings for drive beams~\cite{https://doi.org/10.48550/arxiv.2207.13849, PhysRevLett.128.174803} (both lasers and particle beams) that provide the highest quality self-inject beam or efficiency. 
Another area of interest is to determine optimal conditions when there is statistical uncertainty in the problem, such as shot-to-shot changes to the density profile and beam distribution functions.

\section*{Acknowledgments}
This material is based upon work supported by CAMPA and ComPASS-4, projects of the U.S. Department of Energy, Office of Science, Office of Advanced Scientific Computing Research and Office of High Energy Physics, Scientific Discovery through Advanced Computing (SciDAC) program
under Contract Nos.\ DE-AC02-06CH11357 and DE-AC02-05CH11231, and through FNAL subcontract 644405 and DOE HEP grant DE SC0010064, and NSF award 2108970.
We gratefully acknowledge the computing resources provided on Bebop, a high-performance computing cluster operated by the Laboratory Computing Resource Center at Argonne National Laboratory.

\nocite{*}
\bibliography{./bibs/ref}  

\framebox{\parbox{.90\linewidth}{\scriptsize The submitted manuscript has been created by UChicago Argonne, LLC, Operator of Argonne National Laboratory (``Argonne''). Argonne, a U.S.\ Department of Energy Office of Science laboratory, is operated under Contract No.\ DE-AC02-06CH11357.  The U.S.\ Government retains for itself, and others acting on its behalf, a paid-up nonexclusive, irrevocable worldwide license in said article to reproduce, prepare derivative works, distribute copies to the public, and perform publicly and display publicly, by or on behalf of the Government.  The Department of Energy will provide public access to these results of federally sponsored research in accordance with the DOE Public Access Plan \url{http://energy.gov/downloads/doe-public-access-plan}.}}

\end{document}